%% file: main.tex
\documentclass[
11pt, % The default document font size, options: 10pt, 11pt, 12pt
oneside, % Two side (alternating margins) for binding by default, uncomment to switch to one side
english, % other languages available
singlespacing, % Single line spacing, alternatives: onehalfspacing or doublespacing
]{McMasterThesis} % The class file specifying the document structure

\usepackage[utf8]{inputenc} % Required for inputting international characters
\usepackage[T1]{fontenc} % Output font encoding for international characters

\usepackage{lmodern} % could change font type by calling a different package
\usepackage{lastpage} % count pages
\usepackage{siunitx} % for scientific units (micro-liter, etc)
 \usepackage{setspace}
 \usepackage[utf8]{inputenc}
\DeclareUnicodeCharacter{2212}{-}
\usepackage[backend=biber, giveninits=true, doi=false, natbib=true, url=false, eprint=false, style=authoryear, sorting=nyt, maxcitenames=2, maxbibnames=99, uniquename=false, uniquelist=false, dashed=false]{biblatex} % can change the maxbibnames to cut long author lists to specified length followed by et al., currently set to 99.
\DeclareFieldFormat[article,inbook,incollection,inproceedings,patent,thesis,unpublished]{title}{#1\isdot} % removes quotes around title
\renewbibmacro*{volume+number+eid}{%
  \printfield{volume}%
  \printfield{number}%
  \setunit{\space}%
  \printfield{eid}}
\DeclareFieldFormat[article]{number}{\mkbibparens{#1}}
\renewbibmacro{in:}{\ifentrytype{article}{}{\printtext{\bibstring{in}\intitlepunct}}} % this remove the "In: Journal Name" from articles in the bibliography, which happens with the ynt 
\renewbibmacro*{note+pages}{%
    \printfield{note}%
    \setunit{,\space}% could add punctuation here for after volume
    \printfield{pages}%
    \newunit}    
\DefineBibliographyStrings{english}{% clears the pp from pages
  page = {\ifbibliography{}{\adddot}},
  pages = {\ifbibliography{}{\adddot}},
} 
\DeclareNameAlias{sortname}{family-given}
\usepackage[autostyle=true]{csquotes} % Required to generate language-dependent quotes in the bibliography
\usepackage{colortbl}
\usepackage{babel}

\usepackage{pdflscape}
\newcommand\fh{\babelhyphen{hard}}

\usepackage{parskip} %this will put spaces between paragraphs
\usepackage{acro}
\DeclareAcronym{est}{
	short = EST,
	long  = expressed sequence tags
}

\DeclareAcronym{Xl}{
	short = \textit{X.~laevis},
	long  = \textit{Xenopus~laevis}
}
\DeclareAcronym{Xg}{
	short = \textit{X.~gilli},
	long  = \textit{Xenopus~gilli}
}

\usepackage{etoolbox}
\preto\chapter{\acresetall} % resets acronyms for each chapter

\usepackage{xspace} %helps spacing with custom commands. 

\usepackage{pgfplotstable} % a much better way to handle tables
\pgfplotsset{compat=1.12}

\thesistitle{Quantum Deep Dreaming: A Novel Approach for Quantum Circuit Design} % Your thesis title, print it elsewhere with \ttitle
\supervisor{Dr. Alán \textsc{Aspuru-Guzik}} % Your supervisor's name, print it elsewhere with \supname
\examiner{} % Your examiner's name, print it elsewhere with \examname
\degree{Undergraduate Physics Thesis} % Your degree name, print it elsewhere with \degreename
\author{Romi \textsc{Lifshitz}} % Your name, print it elsewhere with \authorname
\addresses{} % Your address, print it elsewhere with \addressname

\subject{Physics} % Your subject area, print it elsewhere with \subjectname
\keywords{} % Keywords for your thesis, print it elsewhere with \keywordnames
\university{\href{http://www.mcmaster.ca/}{McMaster University}} % Your university's name and URL, print it elsewhere with \univname
\department{\href{https://physics.mcmaster.ca/}{Department of Physics and Astronomy}} % Your department's name and URL, print it elsewhere with \deptname
\group{\href{https://www.matter.toronto.edu/}{The Matter Lab}} % Your research group's name and URL, print it elsewhere with \groupname
\faculty{\href{https://artsci.mcmaster.ca/}{Arts and Science Program}} % Your faculty's name and URL, print it elsewhere with \facname

\hypersetup{pdftitle=\ttitle} % Set the PDF's title to your title
\hypersetup{pdfauthor=\authorname} % Set the PDF's author to your name
\hypersetup{pdfkeywords=\keywordnames} % Set the PDF's keywords to your keywords

\begin{document}
 \frontmatter % Use roman page numbering style (i, ii, iii, iv...) for the pre-content pages

\pagestyle{plain} % Default to the plain heading style until the thesis style is called for the body content

%----------------------------------------------------------------------------------------
%	Half Title (lay title)
%----------------------------------------------------------------------------------------
%\begin{halftitle} % could not get this environment working
%\vspace*{\fill}
\vspace{6cm}
\begin{center}
Quantum Deep Dreaming: A Novel Approach for Quantum Circuit Design % ideally, but it doesn't seem to matter
\end{center}
%\vspace*{\fill}
\pagenumbering{gobble} % leave this here, McMaster doesn't want this page numbered
%\end{halftitle}
\clearpage

%----------------------------------------------------------------------------------------
%	TITLE PAGE
%----------------------------------------------------------------------------------------
\pagenumbering{gobble}
\begin{center}

\vfill
\textsc{\Large \ttitle} \\

\vfill
By \authorname, \\%% -----> List prior degrees after comma  <----

 \vfill
{\large \textit{A Thesis Submitted to the School of Undergraduate Studies in the Partial Fulfillment of the course PHYSICS 4P06: \degreename}}\\

\vfill
{\large \univname\, \copyright\, Copyright by \authorname\, April 19, 2022}\\[4cm] % replace \today with the submission date

\end{center}

%----------------------------------------------------------------------------------------
%	Descriptive note numbered ii
%----------------------------------------------------------------------------------------
% Need to add below info
\newpage
\pagenumbering{roman} % leave to turn numbering back on
\setcounter{page}{2} % leave here to make this page numbered ii, a Grad School requirement

\noindent % stops indent on next line
\univname \\ 
\degreename\, (\the\year) \\
Hamilton, Ontario (\deptname) \\[1.5cm]
TITLE: \ttitle \\
AUTHOR: \authorname\,  %list previous degrees
(\univname)  \\
SUPERVISOR: \supname\, \\ 
NUMBER OF PAGES: \pageref{lastoffront}, \pageref{LastPage}  % put in iv and number

\clearpage

%----------------------------------------------------------------------------------------
%	Lay abstract number iii
%----------------------------------------------------------------------------------------
% not actually included in most theses, though requested by the GSA
% uncomment below lines if you want to include one
%\section*{Lay Abstract}
%\addchaptertocentry{Lay Abstract}
% Type it here
%\clearpage
%----------------------------------------------------------------------------------------
%	ABSTRACT PAGE
%----------------------------------------------------------------------------------------

\section*{\Huge Abstract} 
\addchaptertocentry{Abstract}
% Type your abstract here. 
One of the challenges currently facing the quantum computing community is the design of quantum circuits which can efficiently run on near-term quantum computers, known as the quantum compiling problem. Algorithms such as the Variational Quantum Eigensolver (VQE), Quantum Approximate Optimization Algorithm (QAOA), and Quantum Architecture Search (QAS) have been shown to generate or find optimal near-term quantum circuits. However, these methods are computationally expensive and yield little insight into the circuit design process. In this paper, we propose Quantum Deep Dreaming (QDD), an algorithm that generates optimal quantum circuit architectures for specified objectives, such as ground state preparation, while providing insight into the circuit design process. In QDD, we first train a neural network to predict some property of a quantum circuit (such as VQE energy). Then, we employ the Deep Dreaming technique on the trained network to iteratively update an initial circuit to achieve a target property value (such as ground state VQE energy). Importantly, this iterative updating allows us to analyze the intermediate circuits of the dreaming process and gain insights into the circuit features that the network is modifying during dreaming. We demonstrate that QDD successfully generates, or ‘dreams’, circuits of six qubits close to ground state energy (Transverse Field Ising Model VQE energy) and that dreaming analysis yields circuit design insights. QDD is designed to optimize circuits with any target property and can be applied to circuit design problems both within and outside of quantum chemistry. Hence, QDD lays the foundation for the future discovery of optimized quantum circuits and for increased interpretability of automated quantum algorithm design.

\clearpage
%----------------------------------------------------------------------------------------
%	ACKNOWLEDGEMENTS
%----------------------------------------------------------------------------------------

\begin{acknowledgements}
\addchaptertocentry{\acknowledgementname} % Add the acknowledgments to the table of contents

To my advisor, Prof. Alán Aspuru-Guzik, for his continuing support and the opportunity to take part in the inextricably amazing field of quantum computing. It is an honour and privilege to be able to take part in the research at the University of Toronto Matter Lab. In addition, this research would not have been possible without the support and collaboration of Dr. Chong Sun, Dr. Jakob Kottmann, Dr. Thi Ha Kyaw, Dr. Sukin Sim, and Abhinav Anand.

To my mother, who encourages me to pursue the best version of myself. To my father who shows me how to be kind to other people. To my brother who shows me how to enjoy life. 

\end{acknowledgements}

%----------------------------------------------------------------------------------------
%	LIST OF CONTENTS/FIGURES/TABLES PAGES
%----------------------------------------------------------------------------------------

\tableofcontents % Prints the main table of contents

\listoffigures % Prints the list of figures

\listoftables % Prints the list of tables

%----------------------------------------------------------------------------------------
%	ABBREVIATIONS
%----------------------------------------------------------------------------------------
% many theses don't use this section, as it will be declared at first use and again each chapter. Uncomment these four lines to activate if you want
%\clearpage
%\section*{\Huge Acronyms}
%\addchaptertocentry{Acronyms}
%\printacronyms[name] % name without an option stops the header

%----------------------------------------------------------------------------------------
%	DECLARATION PAGE
%----------------------------------------------------------------------------------------

\begin{declaration}
\addchaptertocentry{\authorshipname}

\noindent I, \authorname, declare that this thesis titled, \enquote{\ttitle} and the work presented in it are my own. I confirm that:

\begin{itemize} 
\item I wrote every chapter and section of this thesis
\item I conducted the training and dreaming of the Quantum Deep Dreaming models described in this report
\item I conducted all of the experiments in Chapters 3 and 4, as well as gathered, cleaned, reported, and analyzed the results in Chapters 3 and 4
\item I developed the analytic techniques for evaluating the performance of Quantum Deep Dreaming and gaining insights from the Quantum Deep Dreaming process, described and used in Chapters 2.3, 3, and 4
\item I created all figures, images, and tables in this document
\end{itemize}
 
\end{declaration}

%%%%%%%%%%%%%%%%%%%%%%%%%%%
%%%%%%%%%%%%%%%%%%%%%%%%%%%
% optional page stuff
%----------------------------------------------------------------------------------------
% can do physical constraints and symbols pages, see the original thesis example on overleaf if you want to include them at https://www.overleaf.com/latex/templates/template-for-a-masters-slash-doctoral-thesis/mkzrzktcbzfl#.VlPeicorpE4
%----------------------------------------------------------------------------------------

%----------------------------------------------------------------------------------------
%	QUOTATION PAGE
%----------------------------------------------------------------------------------------

%\vspace*{0.2\textheight}

%\noindent\enquote{\itshape Thanks to my solid academic training, today I can write hundreds of words on virtually any topic without possessing a shred of information, which is how I got a good job in journalism.}\bigbreak

%\hfill Dave Barry

%----------------------------------------------------------------------------------------
%	DEDICATION
%----------------------------------------------------------------------------------------

% \dedicatory{For/Dedicated to/To my\ldots} 

%%%%%%%%%%%%%%%%%%%%%%%%%%%
%%%%%%%%%%%%%%%%%%%%%%%%%%%
%%%%%%%%%%%%%%%%%%%%%%%%%%%

%----------------------------------------------------------------------------------------
% The following bit is just here to make sure we end up on a new page and get the total number of roman numeral
\label{lastoffront}
\clearpage
% make sure this command is on the last of your frontmatter pages, i.e. only this command, a \clearpage then \mainmatter
% should be fine without modification
%----------------------------------------------------------------------------------------

%----------------------------------------------------------------------------------------
%	THESIS CONTENT - CHAPTERS
%----------------------------------------------------------------------------------------

\mainmatter % Begin numeric (1,2,3...) page numbering

\pagestyle{thesis} % Return the page headers back to the "thesis" style

% Include the chapters of the thesis as separate files from the Chapters folder
\input{Chapters/Chapt1}

\input{Chapters/Chapt2}
\input{Chapters/Chapt3}
\input{Chapters/Chapt4}
\input{Chapters/Chapt5}

% I suggest only compiling one chapter at a time, and comment out the others. That way, the document will typeset faster. When your done with all the chapters, then uncomment them all. Don't worry about the numbering of chapters/figures/etc. LaTeX will take care of that. 

%----------------------------------------------------------------------------------------
%	THESIS CONTENT - APPENDICES
%----------------------------------------------------------------------------------------

\appendix % Cue to tell LaTeX that the following "chapters" are Appendices
\renewcommand{\thetable}{A\arabic{chapter}.\arabic{table}} % adds an A to table names in appendix (Table A1.1, A1.2...)
\renewcommand{\thefigure}{A\arabic{chapter}.\arabic{figure}} % same for figures
\renewcommand{\thesection}{A\arabic{section}}

% Include the appendices of the thesis as separate files from the Appendices folder
\input{Appendix/Supp_Chap1.tex}

%----------------------------------------------------------------------------------------
%	BIBLIOGRAPHY
%----------------------------------------------------------------------------------------

\printbibliography[heading=bibintoc]

%----------------------------------------------------------------------------------------

\end{document}

%% file: Chapters/Chapt1.tex
% Example chapter, could be the introduction

\chapter{Introduction} % Main chapter title

\label{Introduction} %for referencing this chapter elsewhere, use \ref{Introduction}

\section{Introduction}

Quantum computers harness the properties of quantum mechanical phenomena to perform operations which would otherwise be intractable on today’s classical
computers. As a result, fault-tolerant quantum computers are expected to solve classically intractable problems in quantum chemistry, physics, combinatorial optimization, and beyond \citep{Bharti_2022}. However, noise resulting from decoherence of qubits, the fundamental units of information in quantum computers, has been a major barrier to scalable quantum computation. To mitigate the noise and capitalize on the size of current, noisy quantum computers, various Noisy Intermediate Scale Quantum (NISQ) \citep{Preskill_2018} algorithms have been developed such as the Variational Quantum Eigensolver (VQE) \citep{peruzzo2014variational}, Quantum Approximate Optimization Algorithm (QAOA) \citep{farhi2014quantum}, and Quantum Architecture Search (QAS) \citep{du2020quantum}. These algorithms attempt to find quantum circuits which can be run on near-term quantum computers, particularly for applications in quantum chemistry and combinatorics. 

For any NISQ algorithm, the main obstacle is identifying the optimal parametrized quantum circuit  (ansatz) topology for the desired circuit behaviour. Specifically, an ansatz can be defined as $U(\theta)$, where $U$ is a set of unitary gates with parameters $\theta$, which operate on an $n$ qubit system. In variational quantum algorithms, the ansatz parameters are optimized to satisfy a particular objective function \citep{Bharti_2022}. For example, VQE is a standard approach in the field for performing such optimization. VQE optimizes the parameters of an ansatz to satisfy an objective function. However, both classical and quantum computation are used in tandem: the objective function between an ansatz and an objective are evaluated on a quantum computer, and the result is used as the loss during classical training of the ansatz parameters \citep{peruzzo2014variational}.

While variational quantum algorithms provide us with the parameters for reaching a particular objective, the challenge remains of finding the optimal ansatz topology to begin with \citep{Bharti_2022}. One method that has been proposed for finding a quantum circuit is Quantum Architecture Search (QAS) \citep{du2020quantum}. QAS searches through a dataset of potential circuit topologies and finds ones that are optimal candidates for the task at hand. In combination with VQE, both methods can then be used to optimize the circuit ansatz followed by the ansatz parameters.

Despite this progress, it is still not yet understood how the ansatz should be chosen for desired applications. There is lacking human intuition into the properties that make one ansatz more successful at achieving a target objective over another. While properties such as low circuit depth and number of qubits have been shown to be negatively correlated with noise levels, and hence to improve circuit performance, there has been little advancement in understanding of new circuit properties of the same caliber. Existing circuit design algorithms such as VQE and QAS could potentially be run on a large dataset of circuits and set of objectives to investigate such circuit properties, but this would be computationally expensive and very time consuming. Hence, there is a need for a better tool for investigating and discovering properties of successful quantum circuits.

An existing method within machine learning, Deep Dreaming \citep{mordvintsev2015inceptionism}, has been used to understand the properties of images \citep{mordvintsev2015inceptionism} and molecules \citep{shen2021deep}. More specifically, Deep Dreaming, has been used to iteratively design objects (i.e., images, molecules) to reach a target property. Importantly, Deep Dreaming is a method for visualizing the changes that are made to an initial input object in order to reach the final target property. Deep Dreaming, therefore, sounds like a promising tool for generating quantum circuits (objects) while investigating the properties that enable them to reach particular applications (target properties).

% (also known as inverse design of artificial neural networks)

In this paper, we use the Deep Dreaming machine learning technique to understand the design of quantum circuits for different sets of problems. In particular, we propose Quantum Deep Dreaming (QDD), a new algorithm for generating quantum circuits with
a target property, and which provides insights into the circuit design process. QDD iteratively optimizes an initial quantum circuit towards a desired target property, while aiding in the discovery of properties that enable optimal circuit performance. To our knowledge, QDD is the first use-case of Deep Dreaming in quantum computation. We demonstrate proof-of-concept of the new QDD algorithm for ground state preparation and, in doing so, lay the foundation for the future discovery of optimized quantum circuits and for increased interpretability of automated quantum algorithm design.

\section{Background}
In this section, we provide background on common quantum circuit design techniques as well as the Deep Dreaming machine learning technique.

\subsection{Quantum Circuit Design}

\subsubsection{Variational Algorithms}
Variational Algorithms \citep{Bharti_2022} optimize the parameters of a given circuit ansatz to satisfy a specified objective. Circuit ansatz refers to a quantum circuit's initial architecture (topology), which consists of the gates used in the circuit and their location with respect to each other and the circuit qubits. A given ansatz may have static gates (gates with fixed parameters), or parametrized gates (gates with relaxed, or tune-able, parameters). Given an initial ansatz, a Variational Algorithm tunes the parametrized gates through a hybrid quantum-classical technique. The ansatz is initialized with a set of random parameters, and a quantum computer is used to estimate an observable related to the quantum circut (i.e., ground state energy). The observable is then compared to a target observable in a cost function, and this cost function is fed into a classical optimization algorithm which updates the parameters of the ansatz on a classical computer in a direction which minimizes the cost. The updated ansatz and parameters are then fed back into the quantum computer for additional measurement, and the process is repeated iteratively until the cost is minimized, thereby reaching an ansatz that is optimized for the objective at hand.

Variational Quantum Eigensolvers (VQEs) \citep{peruzzo2014variational} are a commonly used subset of Variational Algorithms which aim at preparing ground state circuits. In VQE, the cost function describes a difference between the measured energy of an ansatz, as measured on a quantum computer, and a known target ground state energy. Through the same iterative process described above, a hybrid quantum-classical approach is used to optimize the parameters of an initial ansatz towards ones that prepare the quantum system in their ground state. The major proposed benefit of designing quantum circuits with VQE is the ability to surpass the exponential complexity blow-up of ground state estimation and preparation techniques for large quantum systems on classical computers. The problem of ground state preparation is of importance to this paper, as we use the task of ground state preparation in later sections to benchmark the newly proposed algorithm.

Despite their popularity, VQEs have several limitations, mainly: (1) reachability and (2) expressibility. First, as the ansatz is fixed from the start of the optimization, it may be the case that we begin with an ansatz that is unable to reach our desired state in Hilbert Space. Such an ansatz is one that has low `reachability'. More precisely, reachability is defined as the ability of a parametrized quantum circuit to represent a quantum state which minimizes an objective function \citep{Bharti_2022, akshay2020reachability}. Second, ``expressibility" is defined as ``whether a PQC [\,parametrized quantum circuit]\, is able to generate a rich class of quantum states" \citep{Bharti_2022, sim2019expressibility}. It is often challenging to design an initial ansatz which is both expressive enough to effectively explore the optimization search space, while compact enough to reduce noise for hardware efficiency \citep{sim2019expressibility}. 

\subsubsection{Quantum Architecture Search Algorithms}
Quantum Architecture Search (QAS) \citep{du2020quantum} is a technique aimed at optimizing a quantum circuit's architecture, rather than a quantum circuit's parameters. More specifically, the goal of QAS is to both optimize the architecture of a quantum circuit for the task at hand while minimizing the negative factors that additional gates contribute to a circuit such as decoherence \citep{stilck2021limitations} and barren plateaus \citep{mcclean2018barren, grant2019initialization}. Given an established subspace of circuit architectures, QAS is able to find optimal architectures that satisfy a particular objective \citep{du2020quantum}. 

To specify the subspace of possible architectures and search for an optimal match, QAS utilizes the concept of a quantum neural network (QNN). The QNN is a central model within quantum machine learning which combines the concept of a neural network in artificial intelligence and quantum circuits in quantum computation. There have been multiple  proposals for the design and implementation of a QNN, as this is a central concept and tool within the field of quantum machine learning \citep{schuld2014quest, kwak2021quantum, alam2022qnet, farhi2014quantum, cong2019quantum}. QAS utilizes a QNN model that assumes the circuit-based model of quantum computation, in which quantum gates can be thought of as analogous to nodes within a classical neural network. The parameters of the gates in a quantum circuit can be loosely thought of as analogous to the weights of a classical neural network. 

Now that we have an understanding of a QNN, we may proceed to describe the QAS process \citep{du2020quantum}. QAS creates a large QNN, coined a `supernetwork', which contains all of the possible gate components and qubit connectivities that we may want to include in our quantum circuit. The supernetwork is then trained for a desired task, and subsets of gates within the trained supernetwork are sampled in order to form a smaller network, a `subnetwork'. The subnetwork inherits the weights of the supernetwork. Note that the subnetwork is essentially a proposed quantum circuit, which contains both a circuit's \textit{architecture} (the set of gates with a particular connectivity) and the circuit's \textit{parameters} (the parameters of the gates in the quantum circuit). This subnetwork is then evaluated via a feed-forward pass to determine whether it is adequate for the desired task at hand. If the subnetwork is adequate, it is retrained once more to update its gate parameters for the desired task, and becomes the proposed quantum circuit (ansatz) that QAS recommends for further parameter optimization with VQE. However, if the subnetwork evaluates to be inadequate, additional subnetworks are sampled from the supernetwork until an optimal subnetwork is found. The definition of an `adequate' subnetwork is subjective to those running the search algorithm.

An alternative to the original QAS is Neural Predictor Based QAS (NPQAS) \citep{Zhang_2021}. As the QAS optimization is run on a quantum computer, NPQAS adapts the method to be a fully classical machine learning technique. Similarly to QAS, NPQAS searches for optimal quantum circuit architectures that satisfy a particular objective. In NPQAS, a neural network is trained in a supervised learning fashion to be a `neural predictor': a neural network which predicts the property of a quantum circuit. Given a quantum circuit representation, the neural predictor can tell us whether the circuit possesses a desired property (i.e., the ground state energy). Hence, the neural predictor can be used to search for optimal quantum circuit architectures within a given dataset of candidate circuit architectures.

As mentioned above, both QAS and NPQAS have been proposed to work in collaboration with VQE, such that QAS first employs rounds of optimization to find an optimal ansatz architecture for the task at hand, and then employs VQE on the optimal ansatz architecture to find the optimal parameters. An obstacle facing this approach is the computational cost of the optimization rounds of QAS and VQE. Further, the efficiency of these algorithms heavily relies on prior knowledge for constructing the subnetwork or the search datasets. For example, in QAS, the ability of the the algorithm to find quantum circuits may be hindered by a design of the subnetwork which lacks the circuit architecture components necessary for the task at hand. These are all consequences of the algorithms focusing on \textit{searching} for quantum circuits, rather than \textit{generating} quantum circuits. 

Moreover, while both algorithms automate the process of finding optimal circuit architectures, they provide little to no insights into the process of designing circuits for the task at hand. For example, with a search efficiency of 1\% in NPQAS \citep{Zhang_2021}, it would be difficult to find enough quantum circuits in order to be able to see trends in the structure of selected circuits. It may be possible to do so if many samples are taken, or if the search dataset or the number of sampled subnetworks is very large, but these approaches would rely on expensive computational costs.

\subsection{Deep Dreaming}
The new algorithm we propose in this paper is based upon the Deep Dreaming algorithm in classical machine learning (``Inceptionism'') introduced by Mordvintsev et al. \citep{mordvintsev2015inceptionism}. Initially, Mordvintsev et al. introduced Deep Dreaming as a method for creating images that display the patterns learned in a specific layer of a Convolutional Neural Network (CNN). In the initial proposal of Deep Dreaming, a CNN was trained to classify images, and then its weights and biases were frozen. Upon freezing the model, an image was passed in a feed-forward fashion through the CNN and the sum of the activations of the neurons at a desired layer was computed–this was the objective function (loss) in dreaming process. Once the loss was computed, its gradients were backpropagated to update each pixel of the input image in the direction that maximized the loss. The updated input image was now a brand new image that is a visualization of the features that the neural network learns are important for classification of the images. This process can be done iteratively: feeding in an input image, updating it, and feeding it back into the model, etc. In this way, a final image can be created which provides us with insights into how the neural network learns to classify images. 

Deep Dreaming models can also be viewed as generative models which design inputs that have a target property. For example, Deep Dreaming has been applied by Shen et al. to design molecules with desired logp values \citep{shen2021deep}. We draw upon these works to expand Deep Dreaming into the realm of quantum circuit design. Rather than images or molecules, we evolve quantum circuits towards a particular objective, as described in the next section.

\clearpage

%% file: Chapters/Chapt2.tex
\chapter{Methods} 
\label{Methods}

\section{The Quantum Deep Dreaming Algorithm}

In this section, we describe how to generate novel quantum circuits with a desired property using the Quantum Deep Dreaming (QDD) process. More specifically, QDD is a process for designing circuits with \textit{any} desired target property that can be modeled using a neural network. Let us now describe the QDD algorithm.

We first train an ANN to accurately predict a property of a quantum circuit (i.e., the energy of the quantum circuit). Then, we define the dreaming objective function as the difference between the predicted property value and a target value for this property (i.e., the ground-state energy). Finally, we use the Deep Dreaming process to iteratively update an initial quantum circuit in order to minimize the objective function (via gradient \textit{descent}). With each update to the quantum circuit, another \textit{intermediate circuit} is generated which is closer to the target property. Once the value of the objective function converges to zero, passing the enhanced quantum circuit into the ANN yields a property prediction close to the target value. Importantly, since the ANN is an accurate predictor of the property, it is likely that the enhanced circuit not only yields a predicted property similar to the target, but that the circuit's \textit{true} property value is close to the target value. Thus, we are able to design novel quantum circuits with target properties.

\begin{figure}[ht]
\centering
\includegraphics[width=0.8\textwidth]{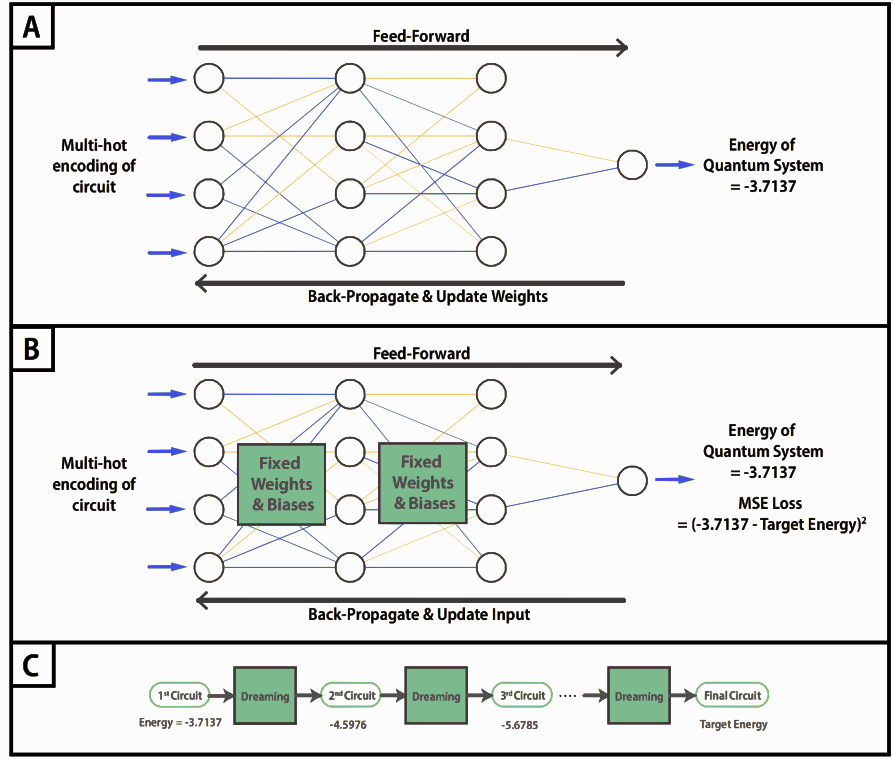}
\caption{Quantum Deep Dreaming. (A) Training, (B) Dreaming, (C) Iterative Dreaming}
\label{fig:qdd}
\end{figure}

\section{Quantum Circuit Representation}
Quantum circuits were represented as multi-hot encoding feature vectors, which we call QCIRCs. One QCIRC was used as the input vector to the neural network during training and dreaming. A new circuit encoding method was used in order to generate the QCIRC representation. The steps of QCIRC encoding are summarized below:

\begin{enumerate}
    \item \textbf{Circuit String:} A circuit string representation is generated, wherein each gate's information is specified by a sequence of characters specifying the gate name, the number of the target qubit, the number of the control qubit (if applicable), and the parameter of the gate (if applicable). More specifically, such a gate sequence is denoted by ``gate=target=control=param". To specify multiple gates in a circuit, gate sequences are concatenated with an ``@" symbol. Note that the pool of gates which can be included in the quantum circuits must also be specified when generating the circuit string.
    \item \textbf{One-Hot and Multi-Hot Encoding:} A one-hot vector encoding of the circuit string is generated. First, each gate sequence in the circuit string is converted into its own one-hot encoded array, after establishing a dictionary for each vector of the gate name, target, control, and parameter. As such, a gate sequence of ``gate=target=control=param" will be convered to a 1x4 array of sub-arrays, where each-sub array indicates the gate name, target, control, and parameter corresponding to the established dictionaries. As there are multiple gate sequences per circuit string, the overall one-hot encoding for a circuit string contains multiple sub-arrays which specify the gate sequences. This overall vector is then converted to a multi-hot encoding, with padding of zeros if necessary.
    \item \textbf{Noise:} Noise between specified lower and upper bounds is randomly added to the features in the multi-hot encoding.  
\end{enumerate}

For example, say that we wish to specify a gate sequence for a CNOT operating on qubit 0 and the target qubit 3 as the control, within a 4-qubit circuit. We first indicate a gate pool of possible gates, such as $P = \{H, X, CRX, CNOT\}$, and then specify the following dictionaries as desired: 

\begin{center}
gate\_dict = \{H:0, X:1, CRX:2, CNOT:3, nop:4\}\\
\smallskip
target\_qubit\_dict = \{Qubit 0:0, Qubit 1:1, Qubit 2:2, Qubit 3:3\}\\
\smallskip
control\_qubit\_dict = \{Qubit 0:0, Qubit 1:1, Qubit 2:2, Qubit 3:3, nop:4\}\\
\smallskip
param\_dict = \{Qubit 0:0, Qubit 1:1, Qubit 2:2, Qubit 3:3, nop:4\}. 
\end{center}

Note that nop is used as a placeholder when no value corresponds to the element in question (i.e., when there is no control qubit or no parameter). Following these specifications, a gate sequence of ``CNOT=0=1=nop'' corresponds to the one-hot encoding,
\begin{center}
[[0, 0, 0, 1, 0], [1, 0, 0, 0], [0, 1, 0, 0, 0], [0, 0, 0, 0, 0]]   
\end{center}
and to the multi-hot encoding,
\begin{center}
[[0, 0, 0, 1, 0, 1, 0, 0, 0, 0, 1, 0, 0, 0, 0, 0, 0, 0, 0]].   
\end{center}
After adding noise with an upper bound of 0.9 and a lower bound of 0.1, we can have a circuit such as,
\begin{center}
[[0.2, 0.1, 1.7, 1.2, 0.6, 1.1, 0.1, 0.4, 0.9, 0.5, 1.2, 0.9, 0.6, 0.8, 0.4, 0.2, 0.1, 0.3, 0.9]].   
\end{center}
A circuit would be a concatenation of such vectors, one vector for each gate sequence, into a larger array. Note that we can also go in reverse to decode a multi-hot encoding of a circuit (the continuous representation) into its circuit string representation (discrete).

\section{Learning from QDD Circuit Design}

Throughout the process of QDD, we have access to the intermediate modified quantum circuits that are generated during the dreaming process. These intermediate circuits provide insight into the sequence of decisions made by the network when designing circuits with the target property. As such, deep dreaming is a powerful tool for both \textit{generating} new inputs that satisfy a certain objective and \textit{learning insights} from the generation process. 

%% file: Chapters/Chapt3.tex
\chapter{Experiments} 
\label{Experiments} 

\section{Experiment Objectives and Success Criteria}
We benchmark the QDD model on the task of ground state preparation. In particular, we use QDD in attempts to generate ground state circuits that fit desired constraints, while providing insight into the process of designing such circuits. To demonstrate proof-of-concept of QDD, we aim to satisfy two criteria: 
\begin{enumerate}
    \item Given an initial circuit with a randomly chosen architecture, is QDD able to learn how to lower the circuit's energy?
    \item Can we interpret how QDD approaches the problem of lowering the energy of quantum circuits towards the ground state?
\end{enumerate}
 Once these two objectives are satisfied, this will show that QDD can assist in the ground state circuit design process and, moreover, assist in interpretable quantum circuit design. In the next section, for the non-quantum chemical audience, we provide background on the problem of ground state preparation and why this problem is meaningful to tackle with QDD.

\section{Primer on Ground State Approximation with Quantum Computation}

The challenge of finding the ground state energy is one that has plagued computational chemistry. To compute the exact ground state energy of a quantum mechanical system, the Hamiltonian operator, $H$, of the system is diagonalized and the minimum eigenvalue is taken to be the ground state energy. However, diagonalization of the Hamiltonian blows-up exponentially with each additional particle that is added to the system. Specifically, for a system of $n$ particles, the Hamiltonian will have dimension $2^n \times 2^n$. Hence, with each addition of a quantum particle to the quantum system, the size of the Hamiltonian increases exponentially, leading to exponential complexity of the classical problem. As such, it is often impossible to exactly determine the ground state energy of large quantum systems in reasonable amounts of time on classical computers. While many techniques have been proposed to approximate the ground state energy for large systems, producing accurate approximations remains a challenge.

Quantum computation is a potential approach to tackling this challenge. By preparing a quantum system and measuring its energy in a quantum computer, rather than simulating the system classically, ground state estimation for large systems may be reduced to one of polynomial time. While it is foreshadowed that an ideal (fault-tolerant) quantum computer will be needed to achieve such runtime, an active area of research is to explore the ability of near-term quantum computers to provide accurate approximations of ground state energy. How is this done, and what are the current obstacles in reaching accurate approximations?

Preparing a ground state on a quantum computer can be reduced to preparing a quantum circuit in its ground state.
Through treating a set of qubits as a quantum chemical system, the system can be evolved towards the ground state using quantum gate operations; note that the set of unitary operations on a quantum system defines a quantum circuit. For example, we can apply an $X$ quantum gate to qubit zero of a quantum system denoted by the wavefunction $\psi$, and then obtain the energy, $E$, of that system through applying the observable $H$:

\[(e^{-i \frac{a}{2} X(q_0)}) (\psi) = \psi_1\] 
\[H \psi_1 = E \psi_1\]    

However, preparing such a quantum circuit requires both hardware and problem-specific considerations. There are a wide variety of quantum computing frameworks that rely on different types of qubits (i.e., photons, spin-based particles, etc.), and which support different native gate sets. Hence, the gates that are used within a quantum circuit must be matched with, or compiled into, ones that are offered by quantum computing hardware. Further, to function on noisy near-term quantum computers, circuits must be designed in ways that minimize the noise that is aggregated when applying operations. To do so, some gates might be preferable over others based on available gate-fidelities, and other properties such as the quantum circuit depth and number of qubits used are desired to be minimized. Beyond hardware, a quantum circuit is specific to the problem at hand. The ground state of two different systems must be reached with different operations, modelling the different hamiltonians of the systems. Hence, new optimal circuits need to be generated for each problem that is approached.

While approaches exist for finding ground state circuits on near-term quantum computers (i.e., VQE), it is still unclear what are the defining properties of ground state circuits, and it is still an obstacle to design ground state circuits which satisfy all of the necessary hardware constraints. We, therefore, use QDD as a tool for overcoming some of these hurdles and assisting in the human design of ground state circuits.

\section{Experimental Setup}
To generate ground state circuits with QDD, we train a model to predict the energy of a circuit and apply deep dreaming on this model to generate circuits close to the ground state energy (the target energy). The model is trained on pairs of QCIRC strings and their corresponding Transverse Field Ising Model (TFIM) energies, computed using a classical simulation of VQE. We choose the TFIM, defined below, as it is commonly used in the literature and allows for comparability with SOTA methods. In this section we will discuss the procedures used to benchmark the QDD algorithm. Section \ref{TFIM} describes how we compute the energy of a quantum circuit. Section \ref{circuit generation} describes the methods used to generate the quantum circuits comprising the training datasets explored.

\subsection{Circuit Energy Computation}
\label{TFIM}
The TFIM Hamiltonian is defined as
$$
    H = -J \left( \sum_{\langle i, j \rangle} Z_iZ_j + g\sum_i X_i \right)
$$
where $J$ is the coupling constant, $g$ is the external field constant, $i$ and $j$ are nearest neighbouring qubits in a quantum circuit, and $X_k$ and $Z_k$ are Pauli matrices applied to a qubit $k \in \{i,j\}$. For a given circuit, we classically simulate its energy using a TFIM with $J=1$, $g=1$, open boundary conditions, and qubits initialized in the zero state, and evaluate the energy using a classical simulation of the VQE using the Tequila \citep{Kottmann_2021} open source package. As the energy landscape of quantum circuits is known to be riddled with local minima, we use three rounds of VQE before assigning a circuit its energy in order to ensure a low energy label. For the rest of this paper, the terms `energy' and `VQE energy' will be used interchangeably as this is how we compute the metric. The described method of energy computation is used to create the labels in all the training datasets. 

\subsection{Dataset Generation}
\label{circuit generation}
We generate multiple datasets with different VQE energy distributions and test the performance of QDD for ground state preparation on these datasets. Initially, we began testing QDD with three datasets that have broad energy distributions. Once optimizing QDD models for these datasets and showing proof-of-concept (described in Section \ref{results}), we tried using datasets with lower and more targeted energy distributions to dream circuits with even lower energies. In the following sections, we will discuss the details of the datasets generated under the `Broad Energy Distribution' and `Targeted Energy Distribution' approaches.

\subsubsection{Broad Energy Distribution Approach}
Three different pipelines are used to generate three datasets with broad energy distributions. Each dataset contains the following:
\begin{itemize}
    \item 5000 randomly generated 6-qubit circuits and their corresponding VQE energies
    \item An equal distribution of 4, 5, 6, 7, and 8 moment circuits
    \item A gate pool of 1-qubit gates (X, Y, Z, H, RX, RY, RZ) and 2-qubit gates (CNOT, CRX, CRY, and CRZ) 
\end{itemize}
Note that since we are using classical simulations for proof-of-concept, we generate only low-depth circuits. We also design our gate pool using insights from quantum chemistry, limiting it to gates that correspond to the Ising Model Hamiltonian and to real eigenstates.

To understand the differences between the datasets, we first note two essential components of any quantum circuit: a circuit's \textit{architecture} (the quantum gates and their locations in the circuit), and  \textit{parameters} (the parameters of the circuit's parametrized quantum gates). A QCIRC string contains information both about a circuit's architecture and parameters. However, when generating a QCIRC string for the three broad energy datasets, only the circuit architecture is specified and the parameters are left unassigned (denoted by `nop' in the parameter component of the circuit string). As the circuit parameters are unassigned and are needed when computing the energy of a circuit, the parameters are randomly initialized when computing the energy.

Now, the differences between the three datasets are the sub-circuits that are used to build each QCIRC string. We use different combinations of the following three sub-circuits to generate a circuit:

\begin{enumerate}
    \item \textbf{Random circuit:} a circuit that is randomly generated using the above specifications, with arbitrarily initialized parameters for the VQE computation.
    \item \textbf{Relaxed mean field (MF) circuit:} the architecture of the circuit that prepares a quantum system in its mean field energy (using only `nop' for the parameter component of parametrized quantum gates). For a 6-qubit system, this circuit is: RY=0=nop=nop0@RY=1=nop=nop1@RY=2=nop=nop2@RY=3=nop=nop3@R\fh Y=4=nop=nop4@R Y=5=nop=nop5. Note that we define the mean field energy and the reasoning for its use below. 
    \item \textbf{Fixed mean field (MF) circuit:} a relaxed mean field circuit with specified, or fixed, parameters. For a 6-qubit system, such a circuit is: RY=0=nop=0.876638666\fh 6903253@RY=1=nop=0.587783873106211@RY=2=nop=0.5334 355932535123@R\fh Y=3=nop=0.5334355932535123@RY=4=nop=0.5877838731062109@RY=5=nop=\fh 0.8766386666903251.
\end{enumerate}

We define the `mean field circuit' as the product ansatz that has the lowest variational energy expectation value that we know of for our 6-qubit system. The energy of this product ansatz is the `mean field energy', and is the estimation of the next lowest energy above the ground state. Hence, the mean field energy is a benchmark for quantifying whether QDD can go beyond this SOTA method of approximating the ground state energy. For reference, note that the mean field energy for our system is -6.902497 and the ground state energy is -7.296230. 

\begin{figure}[ht]
\centering
\includegraphics[scale=0.4]{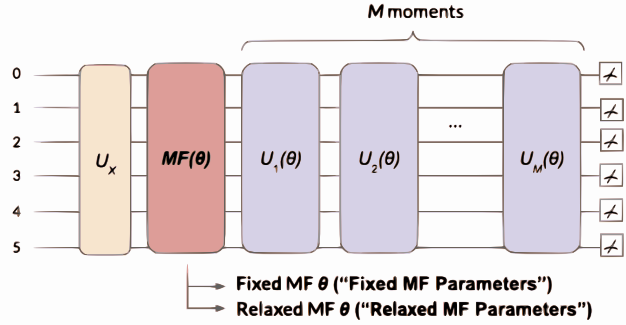}
\caption[Circuit Building Blocks in Datasets $A_s$, $B_s$, $C_s$.]{Circuit Building Blocks in Datasets $A_s$, $B_s$, $C_s$. Circuits in Dataset $A_s$ contain the yellow and purple layers above. Circuits in Datasets $B_s$ and $C_s$ contain all layers above, where the mean field circuit (red) has relaxed parameters and fixed parameters, respectively.}
\label{fig:circuits_mf}
\end{figure}

Finally, using the sub-circuits, we create the three datasets of circuit strings (Figure \ref{fig:circuits_mf}): one dataset with just random circuits (Dataset $A_s$: Without MF), another with the relaxed mean field circuit concatenated to random circuits (Dataset $B_s$: With Relaxed MF), and a final one with the fixed mean field circuit concatenated to random circuits (Dataset $C_s$: With Fixed MF). Note that the mean field circuit was concatenated in order to focus the efforts of dreaming to explore areas beyond the SOTA method of mean field energy computation.

\begin{figure}[ht]
\centering
\includegraphics[scale=0.35]{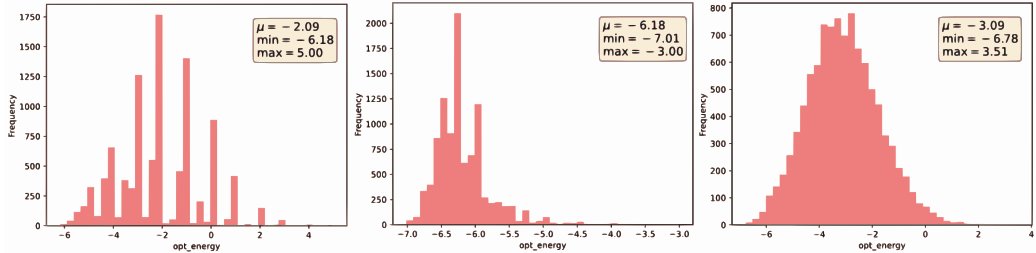}
\caption[Energy Distribution of Datasets $A_s$, $B_s$, $C_s$]{Energy Distribution of Datasets $A_s$, $B_s$, $C_s$ (left to right)}
\label{fig:all_datasets}
\end{figure}

Upon computing the VQE energy for the circuits in each dataset, we now have three datasets prepared in order to train three different QDD models. The energy distribution for each dataset can be seen in Figure \ref{fig:all_datasets}. 

\begin{figure}[ht]
\centering
\includegraphics[scale=0.45]{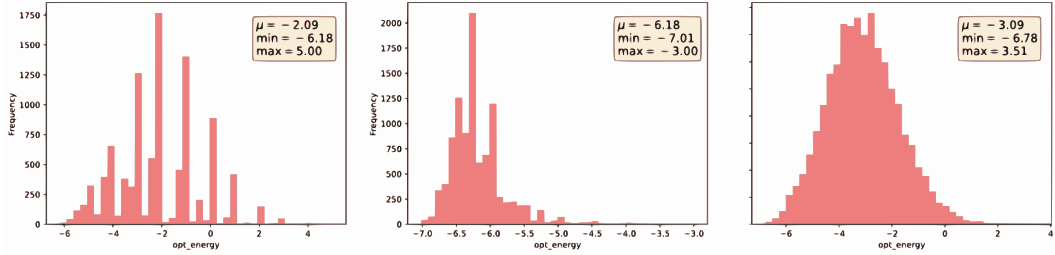}
\caption[Energy Distribution of Datasets $A_l$, $B_l$, $C_l$.]{Energy Distribution of Datasets $A_l$, $B_l$, $C_l$ (left to right)}
\label{fig:10000dataset-distrib}
\end{figure}

We also generate three additional 10,000 circuit datasets ($A_l$, $B_l$, $C_l$) to grow the 5000 circuit datasets $A_s$, $B_s$, and $C_s$. This will allow us to use the optimal training parameters found on datasets $A_s$, $B_s$, and $C_s$ on the larger datasets, and to evaluate whether a larger training dataset improves QDD performance. Rather than completely generating new circuits, the 5000 existing circuits were recycled, and their building blocks were rearranged to generate 5000 additional ones. For example, the random circuit component of circuits from the relaxed MF  dataset (size = 5000) was concatenated with a fixed MF circuit to yield an additional 5000-circuit fixed MF dataset. Together with the already existing fixed MF dataset (Dataset $C_s$, size = 5000), this sums to 10,000 fixed MF circuits. A similar method was used to generate the additional 5000 without MF and relaxed MF circuits. Therefore, three new 10,000 circuit datasets were generated: Dataset $A_l$ (Without MF), Dataset $B_l$ (With Relaxed MF), Dataset $C_l$ (With Fixed MF). Their energy distributions can be seen in Figure \ref{fig:10000dataset-distrib}.

\subsubsection{Targeted Energy Distribution Approach}
Two training datasets were generated with a more targeted gate pool in order to observe if distributions concentrated around lower energies can improve the performance of QDD. They were also generated with a simplified gate pool in order to assist in interpreting the properties that QDD is modifying in an initial circuit to attain a lower energy circuit. However, both datasets still use QCIRC to represent 6-qubit quantum circuits, and the method discussed above for computing the energy of a quantum circuit. Since the energies of these datasets were already very low, mean field circuits were not concatenated to the random circuits generated.

\begin{figure}[ht]
\centering
\includegraphics[scale=0.35]{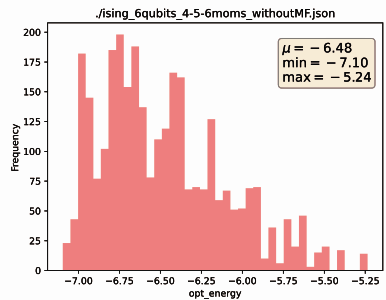}
\caption{Energy Distribution of the RY-CNOT Dataset}
\label{fig:ry-cnot-dataset-distrib}
\end{figure}

The RY-CNOT Dataset uses a gate pool of only \{RY, CNOT\} and arbitrary connectivity. The maximum distance between the target and control qubits is 1 and the ratio of identity to 1 qubit-gates to 2 qubit-gates is [0.2, 0.4, 0.4]. The dataset consists of 3000 circuits total, of which 1000 are 4, 5, and 6 moments each. As can be seen in Figure \ref{fig:ry-cnot-dataset-distrib}, the minimum (-7.10) and maximum (-5.24) energy of this dataset is lower than any of the prior datasets. 

\begin{figure}[ht]
\centering
\includegraphics[scale=0.35]{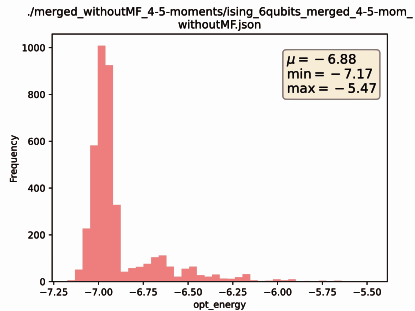}
\caption{Energy Distribution of the XY-Y Dataset}
\label{fig:xy-y-dataset-distrib}
\end{figure}

Lastly, the XY-Y Dataset was generated using a gate pool of only \{XY, Y\}. All-to-all connectivity was used in order to lower the energy of the circuits. For this particular dataset, the circuits were generated in the Tequila framework representation \citep{Kottmann_2021} using circuit generation code from a public GitHub repository \citep{Kottmann2021}. Once generated in Tequila, the circuits were then compiled into the QCIRC representation. This dataset consists of 2000 circuits total, of which 1000 are 4 and 5 moments each. As can be seen in Figure \ref{fig:xy-y-dataset-distrib}, the minimum (-7.17) and maximum (-5.47) energy of this dataset are even lower than the RY-CNOT dataset bounds, and the dataset is very concentrated around an energy of -7.

\section{Training and Dreaming}
Upon generating the datasets, models were trained on each via a supervised learning fashion. Extensive hyperparameter and architecture searches were conducted when training on datasets $A_s$, $B_s$, and $C_s$.

To train a model, first a search was conducted for the architecture, learning rate, and model noise levels. All models for this initial search used the L2 loss ((\textit{Predicted Energy} - \textit{True Energy}$)^2$), no regularization, and the Adam optimizer \citep{kingma2014adam}. PyTorch was the ML framework used \citep{NEURIPS2019_9015}. An 85:15 train:test split was used.               

The search was conducted systematically as described below, and the search settings and results can be found in Appendix \ref{Supp_chap3}. Note that the ‘best’ model in a round is the one which had the lowest test-set loss.
\begin{itemize}
    \item Round 1 - Architecture: The number of nodes in three-layer, fully connected neural network architectures was varied while all other parameters were held constant.
    \item Round 2 - Learning Rate: The learning rate of the best model from round 1 was varied, and all other parameters were held constant.
    \item Round 3 - Noise: The upper-bound of the noise level of the best model from round 2 was varied. The best model from this round was considered the final model of the hyperparameter search.
\end{itemize}

The dreaming performance of the best model found for a dataset was then assessed. To do so, a sample of $n$ circuits was taken from the training dataset and the trained model was used to dream each circuit towards a target energy. The target energy was set to be -8, and an L2 loss between the target energy and the predicted energy of a circuit was used to evaluate whether a circuit was close to the target energy. We refer to this loss as the `dreaming loss', to distinguish it from the loss used in training. All of the intermediate circuits generated during dreaming were recorded, along with their true VQE energies, and the dreaming loss after each intermediate circuit generation. A model's dreaming performance was evaluated using the metrics discussed in Section \ref{Dreaming Evaluation Metrics}. 

Further, to improve the models, we conducted a search for the optimal loss (L1 vs L2) and L2 regularization (weight decay). To do so, for each dataset, the best loss and L2 regularization weight decay was varied systematically to train multiple additional models. The models were then all used to dream $n$ circuits from the training dataset, and the dreaming performance of the models was compared. The tested settings and results can be found in Appendix \ref{Supp_chap3}. Note that for models with no regularization, the Adam optimizer was used. For models with regularization, the AdamW optimizer was used as it has been shown to incorporate L2 regularization more effectively into the optimization \citep{loshchilov2017decoupled}.

For the rest of the datasets, the best hyperparameters and losses found in the extensive hyperparameter search above were carried over. That is, if a dataset was without MF, then the hyperparameters and losses of the best model trained on dataset $A_s$ were used. This was similarly done for the with relaxed MF and with fixed MF datasets.

\section{Dreaming Evaluation Metrics}
\label{Dreaming Evaluation Metrics}
We used multiple metrics to compare between the results of different models, and to ultimately evaluate the success of QDD in satisfying criteria (1) and (2) of the Objectives section. For a trained model that is used to iteratively transform (dream) \textit{one} circuit towards a lower energy circuit, we compute the:
\begin{itemize}
    \item{\textbf{Initial Energy:}} The energy of the initial circuit that is fed into the QDD model.
    \item{\textbf{Final Energy:}} The energy of the final circuit that is generated during dreaming.
    \item{\textbf{Minimum Energy:}} The energy of the lowest energy circuit generated during the dreaming. Note that the minimum energy can correspond to any of the intermediate circuits that were generated during dreaming, inclusive of the initial and final circuits.
    \item{\textbf{Energy Displacement:}} \textit{Final Energy} $-$ \textit{Initial Energy}. 
\end{itemize}

Beyond dreaming one circuit, a QDD model's dreaming performance is evaluated based on the outcome of dreaming multiple circuits. For a sample of \textit{n} circuits that each go through the iterative dreaming process with a given model, we compute the following to gauge a model's `dreaming efficiency':

\begin{itemize}
    \item{\textbf{Percent Final Below Mean Field:}} The percentage of tested circuits that reached a final circuit energy below the mean field energy.
    \item{\textbf{Percent Minimum Below Mean Field:}} The percentage of tested circuits that reached a minimum circuit energy below the mean field energy. 
    \item{\textbf{Lowest Minimum Energy:}} The lowest minimum energy that was attained during the \textit{n} circuits' dreaming processes.
\end{itemize}

It is important to note that while testing dreaming on different models, we were looking for the ability of a model to generate circuits below the mean field energy and hence are computing many of the metrics as relative to the mean field.

\section{Results and Analysis}
\label{results}

\subsection{Broad Energy Distribution Approach}
\subsubsection{Dataset \texorpdfstring{$A_s$}: Without Mean Field}

Upon an extensive hyperparameter and loss search, the optimal model for dataset $A_s$ was found to have a fully-connected network with three hidden layers with 600 neurons each. We used the following hyperparameters: a learning rate of $10^{-6}$, 0.95 upperbound noise level, L1 loss, and L2 regularization with a weight decay of 1. The training and test losses of this model are 2.897 and 2.849, respectively. The losses of the models tested during the hyperparameter search and the dreaming results during the loss search can be seen in Appendix \ref{Supp_chap3}.

The optimal model was used to generate, or dream, 100 quantum circuits. 0\% of tested circuits had a final circuit energy below the mean field (0\% percent final dreamed below mean field). However, 3\% of the circuits tested had a minimum circuit energy below the mean field (3\% percent minimum dreamed below mean field). The lowest final circuit energy was -6.746003281 and the lowest minimum circuit energy was -6.968854603. For dreaming, we used a learning rate of 0.01, 0.9 upperbound noise level, and L2 loss with no L2 regularization.

\subsubsection{Dataset \texorpdfstring{$B_s$}: With Relaxed Mean Field}

Upon an extensive hyperparameter and loss search, the optimal model for dataset $B_s$ was found to have a fully-connected network with three hidden layers, each with 700 neurons. We used the following hyperparameters: a learning rate of $10^{−5}$, 0.95 upperbound noise level, L1 loss, and L2 regularization with a weight decay of 0.0001. The training and test losses of this model are 0.129 and 0.156, respectively. The losses of the models tested during the hyperparameter search and the dreaming results during the loss search can be seen in Appendix \ref{Supp_chap3}.

The optimal model was used to dream 100 quantum circuits. 41\% of tested circuits had a final circuit energy below the mean field (41\% percent final dreamed below mean field). However, 59\% of the circuits tested had a minimum circuit energy below the mean field (59\% percent minimum dreamed below mean field). The lowest final circuit energy was -6.995487124 and the lowest minimum circuit energy was -7.069693743. For dreaming, we used a learning rate of 0.01, 0.9 upperbound noise level, and L2 loss with no L2 regularization. 

\subsubsection{Dataset \texorpdfstring{$C_s$}: With Fixed Mean Field}
Upon an extensive hyperparameter and loss search, the optimal model for dataset $C_s$
was found to have a fully-connected network with three hidden layers with 700 neurons each. We used the following hyperparameters: a learning rate of $10^{−6}$, 0.95 upperbound noise level, L2 loss, and L2 regularization with a weight decay of 0.0001. The training and test losses of this model are 1.657 and 1.836, respectively. The losses of the models tested during the hyperparameter search and the dreaming results during the loss search can be seen in Appendix \ref{Supp_chap3}.

The optimal model was used to generate, or dream, 100 quantum circuits. 4\% of tested circuits had a final circuit energy below the mean field (4\% percent final dreamed below mean field). However, 6\% of the circuits tested had a minimum circuit energy below the mean field (6\% percent minimum dreamed below mean field). The lowest final circuit energy was -7.059580421 and the lowest minimum circuit energy was -7.070890243. For dreaming, we used a learning rate of 0.01, 0.9 upperbound noise level, and L2 loss with no L2 regularization.

\begin{table}[ht]
\centering
\caption{Training Hyperparameter Search and Loss Results: Best Models for Each Dataset (n=5000)}
\label{tab:hyperparam-summary}
\resizebox{\textwidth}{!}{%
\begin{tabular}{|c|c|c|c|c|c|c|} 
\hline
{\textbf{Dataset }} & \multicolumn{3}{c|}{\textbf{Best Training Hyperparameters}}                & {\textbf{Loss}} & {\textbf{Weight Decay}} & {\textbf{Best Test Loss}}  \\ 
\cline{2-4}
                                   & \textbf{Architecture} & \textbf{Learning Rate} & \textbf{Noise Upperbound} &                                &                                        &                                           \\ 
\hline
\textbf{A. Without MF}             & 600, 600, 600         & 1.0e-06                & 0.95                      & L1                             & 1                                      & 2.849\textbf{}                        \\ 
\hline
\textbf{B. With Relaxed MF}        & 700, 700, 700         & 1.0e-05                & 0.95                      & L1                             & 0.0001                                 & 0.156\textbf{}                       \\ 
\hline
\textbf{C. With Fixed MF}          & 700, 700, 700         & 1.0e-06                & 0.95                      & L2                             & 0.0001                                 & 1.836\textbf{}                        \\
\hline
\end{tabular}}
\end{table}

\begin{figure}[ht]
\centering
\includegraphics[scale=0.45]{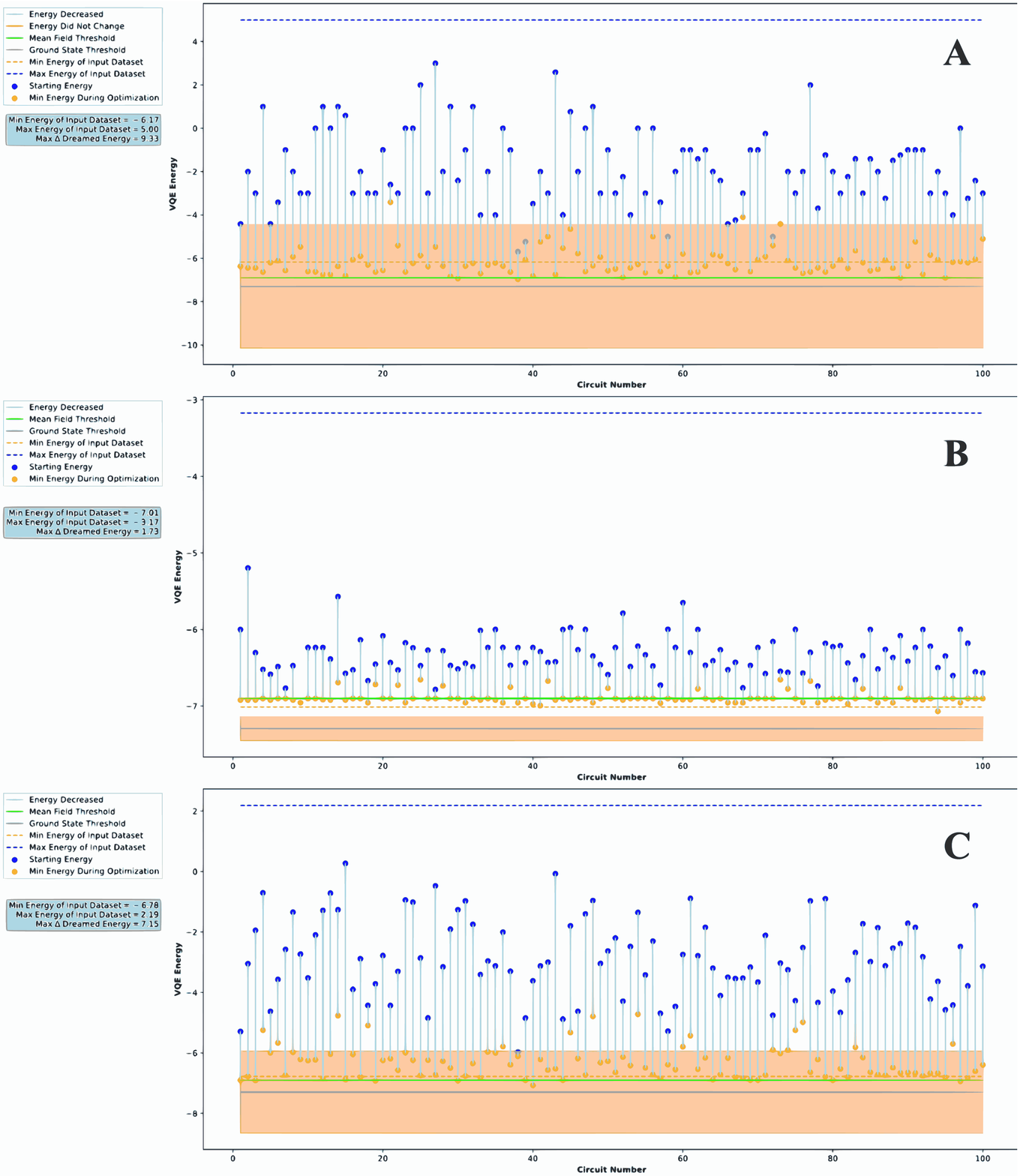}
\caption[Dreaming Progress for Each Circuit Dreamed with Datasets $A_s$, $B_s$, $C_s$]{Dreaming Progress for Each Circuit Dreamed with Datasets $A_s$, $B_s$, and $C_s$. 100 circuits were dreamed using the best model trained on Datasets $A_s$, $B_s$, and $C_s$ (correspondingly labeled). The figures show the start and minimum VQE energy for each dreamed circuit. The orange box encompasses the ground state in the error bounds of the trained model (akin to an error bar). We observe that Datasets $A_s$ and $C_s$ allowed circuits to travel far during dreaming, but to have a small number of circuits pass the mean field energy. Yet, one circuit within Dataset $C_s$ was able to go far beyond the mean field energy. While Dataset $B_s$ allowed circuits to travel smaller distances, the majority of circuits travelled below the mean field energy.}
\label{fig:circuits_mf_2}
\end{figure}

\subsubsection{Datasets \textit{Al, Bl, Cl}}
Recall that datasets $A_l$, $B_l$, and $C_l$ are the same as datasets $A_s$, $B_s$, and $C_s$, respectively, but are simply double in size. Once the optimal models for datasets $A_s$, $B_s$, and $C_s$ were found, the same hyperparameters and losses were used to train three models on the larger datasets $A_l$, $B_l$, and $C_l$. This was done in order to observe whether an increase in the training dataset size could improve the accuracy of the trained models, and hence improve the performance of dreaming. Once trained, the models were used to dream 1000 circuits. In Table \ref{tab:big-small-dreaming-comparison} below, we can observe the differences between the dreaming performance of datasets $A_s$, $B_s$, and $C_s$ and $A_l$, $B_l$, and $C_l$. 

For datasets A and C, we observe that the model achieves a lower testset loss and a better dreaming performance on the larger dataset. However, for dataset B, we observe that the model achieves a lower testset loss but worse dreaming performance. This suggests that a decrease in testset loss may not necessarily result in improved dreaming performance. It should also be noted that the small and large counterparts for each dataset have different training testsets and different dreaming datasets. Therefore, our conclusions for the comparison between the small and large counterparts of each dataset have a degree of uncertainty. Next steps should include comparing between experiments run on identical testsets and dreaming sets between the small and large datasets (i.e., the only change is the size of the trainset).

\begin{table}[ht]
\centering
\caption{Comparison of Training \& Dreaming for Large \& Small Datasets}
\label{tab:big-small-dreaming-comparison}
\resizebox{\textwidth}{!}{%
\begin{tabular}{lcccccc}
                                                                & \multicolumn{2}{c}{\textbf{Without MF}} & \multicolumn{2}{c}{\textbf{With Relaxed MF}} & \multicolumn{2}{c}{\textbf{With Fixed MF}}  \\
                                                                & \textbf{$A_s$} & \textbf{$A_l$}               & \textbf{$B_s$} & \textbf{$B_l$}                    & \textbf{$C_s$} & \textbf{$C_l$}                   \\
\rowcolor[rgb]{0.945,0.945,0.945} \textbf{Training Performance} &             &                           &             &                                &             &                               \\
Trainset Loss                                                   & 2.8971      & 2.6057                    & 0.1293      & 0.1318                         & 1.6573      & 1.6641                        \\
Testset Loss                                                    & 2.8489      & 2.7318                    & 0.1555      & 0.1512                         & 1.8360      & 1.7517                        \\
\rowcolor[rgb]{0.945,0.945,0.945} \textbf{Dreaming Performance} &             &                           &             &                                &             &                               \\
\% Min Dreamed Energy Below MF                                  & 3\%         & 3.80\%                    & 59\%        & 52.10\%                        & 6\%         & 10.80\%                       \\
Lowest Dreamed Energy Achieved                                  & -6.9689     & -6.9971                   & -7.0697     & -7.0437                        & -7.0709     & -7.0372                      
\end{tabular}}
\end{table}

\subsection{Targeted Energy Distribution Approach}
Recall that two additional datasets, RY-CNOT and XY-Y, with lower energy distributions and a simplified gate pool were tested in order to investigate whether QDD models trained on these datasets could reach lower energies. The insights from the hyperparameter, architecture, and loss search from the previous datasets were used to determine initial training settings for models trained on these two datasets. However, it was observed for both datasets that the circuits dreamed were stagnant in energy. That is, dreaming was unable to evolve a circuit towards energies far beyond the initial starting energy of the circuit. Dreaming was able to evolve circuits below the mean field energy, but as the majority of circuits had an initial energy already at the mean field or below, this result holds little significance.

\begin{figure}[t!]
\centering
\includegraphics[scale=0.35]{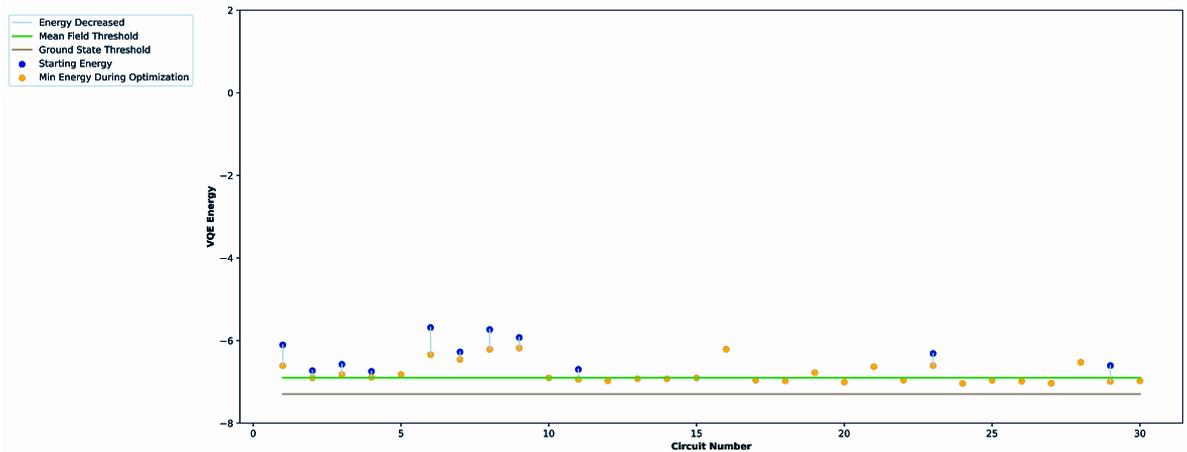}
\caption[Dreaming Progress for Each Circuit Dreamed Using XY-Y or RY-CNOT Datasets.] {Dreaming Progress for Each Circuit Dreamed Using XY-Y or RY-CNOT Datasets. The circuits travelled very short distances and many stayed at their initial energies (the lone orange points). Note that dreaming metrics are not displayed as this figure visualizes circuits dreamed from different models for comparison purposes.}
\label{fig:xy-y-and-ry-cnot-dreamingprog}
\end{figure}

%% file: Chapters/Chapt4.tex
\chapter{Discussion} 
\label{Discussion} 
The primary purpose of this study was to determine whether the QDD algorithm could be used for problem-inspired and quantum resource constraint ansatz design and for interpretability into the ansatz design process. In evaluating the performance of QDD on the first three datasets with broad energy distributions, it is evident that QDD is able to satisfy the proof-of-concept criteria layed out in Section \ref{Experiments}. That is, QDD was able to (1) Lower the energy of given initial circuits towards a target, and (2) Provide visibility and hence, interpretability, into how the model designs quantum circuits with lower energies by showing the intermediate circuits that QDD iterates through to reach the final objective. However, in evaluating the performance of QDD on the datasets with more targeted energy distributions, it is still unclear what is the best methodology for further lowering the energies of the circuits produced by QDD. We will now more closely examine how QDD satisfies the proof-of-concept criteria and potential next steps for reaching lower energy circuits.

\section{Criteria 1: Learning to Lower a Circuit's Energy}
First, across all broad energy distribution datasets, QDD successfully generated new circuits that are closer to the desired target (criteria 1), which in our benchmark test is the ground state energy (-7.29623). To show this, we will focus on the smaller datasets ($A_s$, $B_s$, $C_s$) as both portray similar trends. Specifically, across models $A_s$, $B_s$, and $C_s$, 99.66\% of the circuits tested decreased in energy. Models $A_s$, $B_s$, and $C_s$ (trained on datasets $A_s$, $B_s$, and $C_s$) had an average energy displacement of -4.320, -0.573, and -3.473 units (Table \ref{tab:minEnergy-summary}), and a maximum energy displacement of -9.332, -1.726, and -7.154 units, respectively. Further, all of the models were able to generate circuits with energies below the mean field SOTA method (-6.902497), as model $A_s$, $B_s$, and $C_s$'s minimum energies were -6.968855, -7.069694, and -7.070890, respectively. The dreaming efficiency of transforming a circuit to be below the mean field energy during the dreaming process was highest for model $B_s$ (59\%), followed by model $C_s$ (6\%) and $A_s$ (3\%). All of these results demonstrate that QDD is able to understand the desired task of modifying a circuit towards a desired target property, so long as that target property can be expressed in a differentiable metric that can undergo the process of backpropagation during neural network training and dreaming. 

\begin{table}[ht]
\centering
\caption{Change in Energy During QDD}
\label{tab:minEnergy-summary}
\resizebox{\textwidth}{!}{%
\begin{tabular}{|l|c|c|c|} 
\hline
\textbf{Dataset}      & \begin{tabular}[c]{@{}c@{}}\textbf{Average Initial}\\\textbf{Energy}\end{tabular} & \begin{tabular}[c]{@{}c@{}}\textbf{Average Minimum Energy}\\\textbf{Achieved in a Dreaming Run}\end{tabular} & \begin{tabular}[c]{@{}c@{}}\textbf{Difference in~}\\\textbf{Average Energies}\end{tabular}  \\ 
\hline
\textbf{$A_s$. WithoutMF} & -1.871                                                                             & -6.192                                                                                        & -4.320                              \\ 
\hline
\textbf{$B_s$. RelaxedMF} & -6.321                                                                             & -6.894                                                                                        & -0.573                              \\ 
\hline
\textbf{$C_s$. FixedMF}   & -2.881                                                                             & -6.355                                                                                        & -3.473                              \\
\hline
\end{tabular}}
\end{table}

\begin{table}[ht]
\centering
\caption{Change in Expressibility During QDD}
\label{tab:minExpr-summary}
\resizebox{\textwidth}{!}{%
\begin{tabular}{|l|c|c|c|} 
\hline
\textbf{Dataset}      & \begin{tabular}[c]{@{}c@{}}\textbf{Average Initial}\\\textbf{Express.}\end{tabular} & \begin{tabular}[c]{@{}c@{}}\textbf{Average Express. of Min}\\\textbf{Dreamed Circuit}\end{tabular} & \begin{tabular}[c]{@{}c@{}}\textbf{Difference in~}\\\textbf{Average Express.}\end{tabular}  \\ 
\hline
\textbf{$A_s$. WithoutMF} & 0.664812                                                                            & 2.781802                                                                                       & 2.116990                                                                                \\ 
\hline
\textbf{$B_s$. RelaxedMF} & 3.300879                                                                            & 3.273963                                                                                       & -0.026916                                                                               \\ 
\hline
\textbf{$C_s$. FixedMF}   & 0.581805                                                                            & 2.425125                                                                                       & 1.843319                                                                                \\
\hline
\end{tabular}}
\end{table}

Looking to the circuits generated, we are also able to assess their quality using the metric of `expressibility'. The expressibility of a circuit is its ``ability to generate (pure) states that are well representative of the
Hilbert space'' \citep{sim2019expressibility}. Expressibility is a useful metric for assessing the quality of quantum circuits as expressible circuits are able to explore many states within the Hilbert space. Further, highly expressible circuits have been shown to improve a circuit's ability to reach the ground state. Hence, highly expressible circuits are often desired ansatze for VQE \citep{sim2019expressibility}. Surprisingly, although the majority of the circuits in datasets $A_s$, $B_s$, and $C_s$ had low expressibility, QDD models $A_s$ and $C_s$ were able to greatly increase the expressibility of circuits (from an average expressibility of 0.664812 to 2.781802 for model $A_s$, and 0.581805 to 2.425125 in model $C_s$) (Table \ref{tab:minExpr-summary}). The increase in expressibility observed through QDD shows that QDD may be a good candidate algorithm for generating quantum ansatze for VQE, as QDD designs ansatze that are more capable of reaching the ground state than randomly designed quantum circuits. Further, the increase in expressibility demonstrates that the lower energy circuits generated align with prior literature on some of the properties that ground state circuits are expected to possess. 

Model $B_s$, however, generated circuits with little to no change in expressibility. A possible reason for this observation is that the mean field circuit was already saturating the expressiblity of the circuit, and so there was little room for improvement. This observation aligns with prior literature, which shows that the expressibility of a circuit can indeed reach a saturation level \citep{sim2019expressibility}. It is unclear, however, why the fixed MF dataset improved so significantly in expressibility as it also has a mean field circuit concatenated to its random circuits. The main difference between the fixed and relaxed MF datasets is that one has a mean field circuit with fixed parameters, and the other has a mean field circuit with relaxed parameters, respectively. It would be interesting to conduct additional experiments exploring how the parameters of the mean field circuit affect the way in which QDD designs circuits. One way to investigate this consists of using both the fixed MF and relaxed MF models to dream the same circuit towards one of lower energy, and then compare what changes the models make to arrive at the final circuits. The possibility for such an in-depth analysis is made possible by the intermediate circuits generated by QDD, and will be expanded upon in Section \ref{criteria 2}.

An important observation of the benchmark results also indicates that the QDD models trained are generalizable. That is, they are able to find circuits with energies lower than the minimum energy of the training dataset. In particular, datasets $A_s$, $B_s$, and $C_s$ were able to go 0.79, 0.29, and 0.06 units below their training dataset energy minimum, respectively. This is consistent with the benchmarks of the SOTA method, NPQAS \citep{Zhang_2021}, and is motivating evidence that QDD is learning inherent, general patterns within the training dataset and applying them in dreaming to design novel circuits. All of the above indicates that QDD is able to learn how to modify an initial quantum circuit towards a desired target property (criteria 1).
\begin{figure}[ht]
\centering
\includegraphics[scale=0.5]{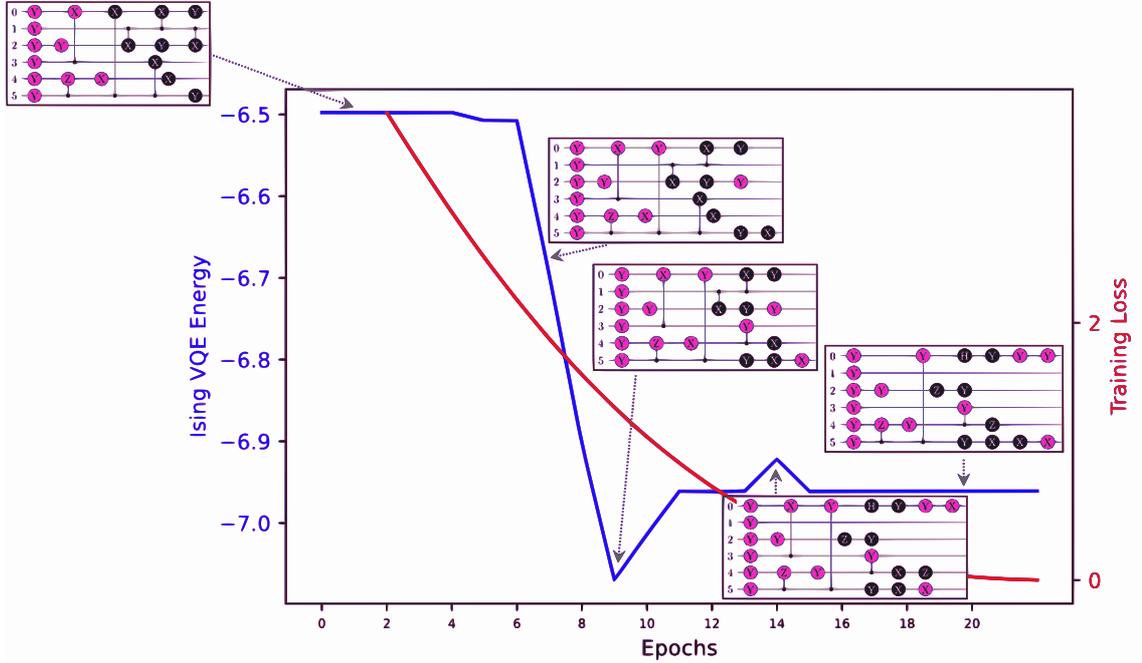}
\caption{In-Depth Dreaming Circuit Generation Analysis for the Best Dreamed Circuit Using Model $B_s$ (Relaxed MF)}
\label{fig:in-depth-dreaming-relaxedMF}
\end{figure}

\section{Criteria 2: Interpretability of Circuit Design}
\label{criteria 2}
Looking to criteria 2, QDD is able to provide rich information about its design process and allow for interpretable quantum circuit design. This is perhaps the most important quality of QDD. Specifically, the intermediate circuit representations throughout the generation process provide an avenue to understanding which properties of the initial circuit are being emphasized or reduced in order to reach the final circuit. To see this more clearly, let us look at an in-depth example.

Figure \ref{fig:in-depth-dreaming-relaxedMF} shows the progression of the intermediate circuits throughout dreaming for the best circuit dreamed using model $C_s$ (relaxed MF dataset). The process begins with an initial randomly generated circuit and evolves towards circuits that are closer to the ground state energy. Focusing firstly on the initial (high energy, epoch 1) and final generated (minimum energy, epoch 9) circuits, we are able to see the changes that a model is making in the circuit in order to reach low energy states. We observe that the circuit’s rotational gates increased from 10 to 14, introducing additional degrees of freedom into the circuit through the rotational parameters. The number of entangling gates was reduced by one, while the symmetry of the all-to-all connectivity within the initial circuit was maintained. The all-to-all connectivity has the potential to result in shorter depth circuits through removing SWAP gates and allowing for long-range entanglement between adjacent 2-qubit quantum gates. So, it is significant that we are able to see QDD maintaining this property throughout the iterative circuit generation. Further, while the number of gates enabling rotations about the Z axis (CRZ, RZ) are maintained, the number of gates enabling rotations about the Y (CRY, RY) and X (CRX, RX) are increased (from 7 to 10, and 2 to 3, respectively). Additionally, the number of two-qubit gates and one-qubit gates decrease and increase, respectively–properties which are important for reducing decoherence on devices \citep{Meier_2003}. Hence, we are able to see that the circuit generated in this case may be a more fitting starting ansatz for VQA algorithms compared to the initially generated random circuit. From this in-depth analysis of the circuit, it is evident that QDD provides a rich source of information on how circuits are designed to achieve certain tasks, and into potential discoveries of properties that characterize classes of quantum circuits. 

It is noteworthy to mention the lack of literature surveying the types of quantum gates, connectivity, and other circuit properties that are important for ground state circuits. Such papers were sought out for in order to better explain whether the evolution of the sample circuit above aligned with existing knowledge of ground state quantum circuits, but only a few papers were found on similar topics \citep{yordanov2020efficient, pfeuty1971ising, whitfield2011simulation, bespalova2021quantum}. This points to the value that QDD brings to the design of such circuits, as QDD allows us to explicitly see the properties modified in circuits to achieve lower energy states. It would be beneficial to conduct an extensive literature search on the topic as a next step in order to discern the extent of the gap in the literature and explore the value that QDD can contribute to the design of low-energy circuits.

\section{Strengths, Limitations, and Next Steps}
A strength of QDD is its capability to \textit{generate} new circuits towards a target property. While existing SOTA methods such as NPQAS \citep{Zhang_2021} and QAS \citep{du2020quantum} are able to \textit{find} circuits in an existing dataset or subspace of possible candidate architectures corresponding to a specified property, QDD is able to generate new fitting circuits given an initial randomly generated candidate architecture. Assuming a high search efficiency for QDD, this means that once the QDD model is trained, QDD only needs to be fed one or a few random circuits which it iterates upon to find a low energy circuit. Such an approach eliminates the need to generate a large dataset of candidate circuits to search within, or to specify a subspace of candidate architectures. 

Additionally, a major strength of QDD is its ability to provide insight and visibility into the circuit design process. In comparison, NPQAS and QAS operate with a black box approach, in which it is difficult to know \textit{why} certain circuits are deemed optimal for the task at hand. Due to QDD's ability to show the intermediate circuits that are iterated upon to achieve a circuit with the desired target property, it is possible to analyze the circuit evolutions and explore any properties that are emphasized or reduced in circuits to achieve the final circuit. Our hope is that, with the proof-of-concept of QDD, in the future QDD can be used to discover new properties of quantum circuits for a variety of objectives. Next steps for the interpretability include simplifying the circuits that are generated. That is, pruning or compiling together any redundant or repetitive gates within the circuit. This may allow us to more easily interpret the properties that QDD is inserting into the circuits to reach lower energies.

Looking to the limitations of QDD, it is noteworthy to mention that while QDD was able to generate low energy circuits comparable to SOTA methods, the models trained were still unable to reach the ground state energy. Looking to the training datasets generated via the broad energy distribution approach, only a small percentage of circuits in these datasets were below the mean field and none were at the ground state. It is likely that circuits in the current datasets are representative of a part of Hilbert space which does not contain the information necessary for higher energy states to reach lower energy states. Hence, a lower energy dataset could be helpful to break into a new subspace of circuits which possess architecture properties from which QDD can learn to reach the ground state energy. Similarly, the targeted energy distribution datasets, albeit possessing circuits below the mean field and concentrated in low energies, did not have any ground state circuits. The targeted energy distributions could be limited by this as well.

Further, an additional plausible explanation is the accuracy of the trained models. The performance of dreaming heavily relies on the training loss of the model. This is due to the fact that, during dreaming, the model that has been trained to predict VQE energy is used in the computation of dreaming loss. That is, recall that the dreaming loss is: (\textit{Predicted Energy} - \textit{Target Energy}$)^2$. If the model is inaccurate (has a high training loss), the predicted energy will be inaccurate and dreaming will have an erroneous intuition into how to evolve a quantum circuit towards the target energy. Indeed, initially, when very inaccurate models were used to dream circuits, the dreaming loss approached zero even when an intermediate circuit's energy was very far from the target energy. With the final models in datasets $A_s, B_s, C_s, A_l, B_l, C_l$, the loss of the models was small and hence the dreaming was more accurate and well-guided. However, these models are still imperfect, as they do not have zero loss, and it was observed that the dreaming loss was still arriving at zero too early. Hence, the models' training losses may be contributing to the fact that energies were unable to reach the ground state, and the training loss appears to be an inherent limitation of QDD. 

By the same logic, we may look to see if a similar trend in dreaming loss is occurring in models trained on the targeted energy distributions. Indeed, in most of the dreaming runs for these models, the dreaming loss was decreasing from one intermediate circuit to the next, despite the energy of the circuits generated being stagnant and far from the target energy. This indicates that the models trained on the targeted energy distributions were inaccurate and misguiding the dreaming. More concretely, this may be happening because the models were trained on a very narrow distribution of energies, and hence, may not be learning the inherent differences between high and low energy circuits. Such insights could be important for the dreaming process, and may explain why the broad distribution datasets exhibited better dreaming performance. However, the reliance of QDD on a diverse dataset requires further exploration, such as testing the performance of QDD on datasets that are targeted around different energy ranges (both high, medium, and low). Further, all of the above shows the inherent limitation of QDD–as it is a supervised learning method, it is limited by the training dataset.

Further, a potential method to remedy both the limitations in the dataset distributions and in the training accuracy is to generate a dataset with even lower energy distributions that reach the ground state energies, alongside inclusion of high energy circuits. Doing so might allow the models to more accurately predict the energies of the lower energy circuits generated during dreaming as well as to learn how to more effectively `travel' from high to low energy circuits. However, generating such a dataset of ground state circuits is a computationally expensive and challenging task in itself. If it was possible to easily curate such a dataset, this would in essence solve a major challenge that quantum computation aims to address. 

Alternatively, another possible direction includes turning to active learning methodologies which continuously feed in the newly generated datasets that are below the training distribution into the model. In doing so, the accuracy of the model might be extended. Note that rather than trying to improve the accuracy during the initial training of the model, this approach shifts the improvement of the accuracy to occur during the dreaming stage.

\section{Applications and Future Directions}
We have, hence, demonstrated a proof-of-concept of the new Quantum Deep Dreaming algorithm and the possibility of using QDD for designing circuits for additional tasks and constraints. In particular, QDD can be an incredibly useful tool for the design of many additional hardware constrained and problem specific circuits. As QDD generates circuits that fit constraint-based specifications (i.e., the gate set, number of qubits, the circuit depth), QDD can also be viewed as a quantum compiling algorithm: an algorithm which finds an optimal quantum circuit for a task given hardware constraints. QDD can easily be specified for desired hardware constraints by creating a training dataset that fits these constraints. For example, in our broad energy distribution benchmark experiment, we limit circuits to be between depth 4-8, and of a particular gate set. The algorithm then learned to generate circuits accordingly. In the NISQ Era where quantum computers are fundamentally limited in traits such as native gate sets, connectivity, and depth optimality, an algorithm such as QDD could be helpful for creating circuits that can run on near-term quantum computers. 

In addition to specifying constraints in the dataset, the objective function of QDD can be used to guide the algorithm to generate circuits for particular hardware constraints. For example, a target property related to noise can be used in order to minimize the noise levels of circuits that are generated. Hence, QDD’s versatility has potential for use in hardware-specific algorithm design. Beyond hardware-specific objectives, QDD has the potential for use to design any problem-inspired task which can be expressed as a differentiable optimization problem.

Beyond quantum computing, QDD has applications within quantum physics and quantum chemistry. For example, a quantum circuit can simply be seen as an idealized simulation of a physical and/or chemical system. Hence, datasets about such systems can be potentially converted into the proposed circuit string representation, and a model can be trained on these datsets and used to dream new physical or chemical systems. Importantly, using a similar analytic process to that described above, this can yield insight into properties that define quantum physical or chemical systems within particular subspaces of the training datasets. So, this has the potential to yield insight into properties that define groups of quantum physical and chemical systems. We have hence proposed a tool that can allow for investigation/interpretability into both quantum computing and multiple other applications within quantum physics and chemistry. 

%% file: Chapters/Chapt5.tex
\chapter{Conclusion} 
\label{Conclusion} 

We have demonstrated proof-of-concept of a new quantum algorithm, Quantum Deep Dreaming, for iterative generation of quantum circuits towards a target property. We benchmarked the algorithm on the task of ground state preparation, and showed that the algorithm is able to learn features within a given training dataset and to utilize these learnings to generate new quantum circuit architectures that progress towards a desired objective. Further, the power of QDD for interpretable quantum algorithm design was demonstrated. Next steps include simplifying the generated circuits for ease of interpretability, exploring active learning methods for further improving QDD performance, and extending the algorithm to generating quantum circuit architectures for both problem-inspired and/or hardware-efficient applications. Additional next steps include integrating QDD into VQA workflows in order to begin optimization with a more expressible, problem-specific, and/or hardware-efficient ansatz. We, therefore, introduce a new algorithm for quantum circut generation that can be used for automated and insightful design of circuit architectures for NISQ algorithms such as VQAs.

%% file: Appendix/Supp_Chap1.tex
\chapter{Chapter 3 Supplement}

\label{Supp_chap3}

Below, we provide the results of the hyperparameter and loss searches for the models trained on datasets $A_s$, $B_s$, and $C_s$. Note that WD stands for Weight Decay. Further, * indicates an optimal model found in a search round, and ** indicates that the model was stopped early during training due to a plateauing loss.

\begin{table}[ht]
\centering
\resizebox{\textwidth}{!}{%
\centering
\begin{tabular}{|l|l|l|l|l|l|l|l|l|} 
\hline
\multicolumn{3}{|c|}{Round 1: Architecture}                                                                                                                                                                                            & \multicolumn{3}{c|}{Round 2: Learning Rate}                                                                                                                                                                                                            & \multicolumn{3}{c|}{Round 3: Noise Upper-bound}                                                                                                                                                                                                                                                       \\ 
\hline
\multicolumn{1}{|c|}{Architecture}  & \multicolumn{1}{c|}{{\begin{tabular}[c]{@{}c@{}}Train \\Loss (L2)\end{tabular}}} & \multicolumn{1}{c|}{{\begin{tabular}[c]{@{}c@{}}Test \\Loss (L2)\end{tabular}}} & \multicolumn{1}{c|}{{Learning Rate}} & \multicolumn{1}{c|}{{\begin{tabular}[c]{@{}c@{}}Train \\Loss (L2)\end{tabular}}} & \multicolumn{1}{c|}{{\begin{tabular}[c]{@{}c@{}}Test \\Loss (L2)\end{tabular}}} & \multicolumn{1}{c|}{{\begin{tabular}[c]{@{}c@{}}Noise \\Upper-bound\end{tabular}}} & \multicolumn{1}{c|}{{\begin{tabular}[c]{@{}c@{}}Train \\Loss (L2)\end{tabular}}} & \multicolumn{1}{c|}{{\begin{tabular}[c]{@{}c@{}}Test \\Loss (L2)\end{tabular}}}  \\
\multicolumn{1}{|c|}{(3 FC layers)} & \multicolumn{1}{c|}{}                                                                           & \multicolumn{1}{c|}{}                                                                          & \multicolumn{1}{c|}{}                               & \multicolumn{1}{c|}{}                                                                           & \multicolumn{1}{c|}{}                                                                          & \multicolumn{1}{c|}{}                                                                             & \multicolumn{1}{c|}{}                                                                           & \multicolumn{1}{c|}{}                                                                           \\ 
\hline
100, 100, 100                       & 2.984                                                                                           & 2.943                                                                                          & e-2                                                 & 3.053                                                                                           & 3.187                                                                                          & 0.9                                                                                               & 2.642                                                                                           & 2.887                                                                                           \\ 
\hline
200, 200, 200                       & 2.907                                                                                           & 2.789                                                                                          & e-3                                                 & 2.624                                                                                           & 3.014                                                                                          & 0.91                                                                                              & 2.532                                                                                           & 3.051                                                                                           \\ 
\hline
300, 300, 300                       & 2.800                                                                                           & 2.877                                                                                          & e-4                                                 & 2.634                                                                                           & 3.012                                                                                          & 0.92                                                                                              & 2.607                                                                                           & 2.757                                                                                           \\ 
\hline
400, 400, 400                       & 2.735                                                                                           & 2.956                                                                                          & e-5                                                 & 2.572                                                                                           & 2.817                                                                                          & 0.93                                                                                              & 2.713                                                                                           & 2.777                                                                                           \\ 
\hline
500, 500, 500                       & 2.762                                                                                           & 2.727                                                                                          & e-6*                                                & 2.707                                                                                           & 2.569                                                                                          & 0.94                                                                                              & 2.633                                                                                           & 2.995                                                                                           \\ 
\hline
600, 600, 600*                      & 2.707                                                                                           & 2.569                                                                                          & e-7                                                 & 2.928                                                                                           & 3.165                                                                                          & 0.95*                                                                                             & 2.707                                                                                           & 2.569                                                                                           \\ 
\hline
700, 700, 700                       & 2.734                                                                                           & 2.648                                                                                          & e-8                                                 & 3.080                                                                                           & 2.979                                                                                          & 0.96                                                                                              & 2.708                                                                                           & 2.869                                                                                           \\ 
\hline
800, 800, 800                       & 2.686                                                                                           & 2.720                                                                                          & e-9**                                               & 6.295                                                                                           & 6.219                                                                                          & 0.97                                                                                              & 2.702                                                                                           & 2.727                                                                                           \\ 
\hline
900, 900, 900                       & 2.692                                                                                           & 2.954                                                                                          & e-10**                                              & 7.499                                                                                           & 7.692                                                                                          & 0.98                                                                                              & 2.655                                                                                           & 2.836                                                                                           \\ 
\hline
1000, 1000, 1000                    & 2.667                                                                                           & 3.172                                                                                          & e-11**                                              & 7.399                                                                                           & 7.312                                                                                          & 0.99                                                                                              & 2.778                                                                                           & 2.889                                                                                           \\ 
\hline
                                    &                                                                                                 &                                                                                                & e-12**                                              & 7.219                                                                                           & 7.272                                                                                          & 1                                                                                                 & 2.745                                                                                           & 2.600                                                                                           \\
\hline
\end{tabular}}
\caption{Dataset $A_s$ Hyperparameter Search}
\end{table}

\begin{table}[ht]
\centering
\resizebox{\textwidth}{!}{%
\centering
\begin{tabular}{|l|l|l|l|l|l|l|l|} 
\hline
\multicolumn{1}{|c|}{{Train Loss}} & \multicolumn{1}{c|}{{Metric (n=100)}} & \multicolumn{6}{c|}{Training Optimizer}                                                                                                                                                                                                       \\ 
\cline{3-8}
\multicolumn{1}{|c|}{}                            & \multicolumn{1}{c|}{}                                & \multicolumn{1}{c|}{{Adam (No Regularization)}} & \multicolumn{5}{c|}{AdamW (L2 Regularization)}                                                                                                                               \\ 
\cline{4-8}
\multicolumn{1}{|c|}{}                            & \multicolumn{1}{c|}{}                                & \multicolumn{1}{c|}{}                                          & \multicolumn{1}{c|}{WD 1}                     & \multicolumn{1}{c|}{WD 0.1} & \multicolumn{1}{c|}{WD 0.01} & \multicolumn{1}{c|}{WD 0.001} & \multicolumn{1}{c|}{WD 0.0001}  \\ 
\hline
\multicolumn{1}{|c|}{L1}                          & \% Final Dreamed Energy Below MF                     & 0\%                                                            & {\cellcolor[rgb]{0.537,0.765,0.537}}0\%       & 1\%                         & 3\%                          & 0\%                           & 0\%                             \\ 
\hline
                                                  & \% Min Dreamed Energy Below MF                       & 2\%                                                            & {\cellcolor[rgb]{0.537,0.765,0.537}}3\%       & 2\%                         & 4\%                          & 1\%                           & 4\%                             \\ 
\hline
                                                  & Lowest Final Dreamed Energy                          & -6.884178                                                      & {\cellcolor[rgb]{0.537,0.765,0.537}}-6.746003 & -6.908385                   & -6.942333                    & -6.767029                     & -6.873292                       \\ 
\hline
                                                  & Lowest Min Dreamed Energy                            & -6.939637                                                      & {\cellcolor[rgb]{0.537,0.765,0.537}}-6.968855 & -6.922309                   & -6.942333                    & -6.917442                     & -6.956455                       \\ 
\hline
                                                  & Training Trainset Loss                               & 2.757                                                          & {\cellcolor[rgb]{0.537,0.765,0.537}}2.897     & 2.753                       & 2.820                        & 2.807                         & 2.706                           \\ 
\hline
                                                  & Training Testset Loss                                & 2.952                                                          & {\cellcolor[rgb]{0.537,0.765,0.537}}2.849     & 2.877                       & 2.529                        & 2.915                         & 3.057                           \\ 
\hline
\multicolumn{1}{|c|}{L2}                          & \% Final Dreamed Energy Below MF                     & 0\%                                                            & 1\%                                           & 1\%                         & 0\%                          & 1\%                           & 0\%                             \\ 
\hline
                                                  & \% Min Dreamed Energy Below MF                       & 2\%                                                            & 3\%                                           & 3\%                         & 0\%                          & 1\%                           & 1\%                             \\ 
\hline
                                                  & Lowest Final Dreamed Energy                          & -6.626004                                                      & -6.956042                                     & -6.917442                   & -6.835520                    & -6.918968                     & -6.902496                       \\ 
\hline
                                                  & Lowest Min Dreamed Energy                            & -6.956456                                                      & -6.956455                                     & -6.956456                   & -6.902496                    & -6.918969                     & -6.902497                       \\ 
\hline
                                                  & Training Trainset Loss                               & 2.694                                                          & 2.706                                         & 2.679                       & 2.678                        & 2.671                         & 2.687                           \\ 
\hline
                                                  & Training Testset Loss                                & 2.866                                                          & 3.048                                         & 3.016                       & 2.957                        & 2.957                         & 2.888                           \\
\hline
\end{tabular}}
\caption{Dataset $A_s$ Loss Search}
\end{table}

\begin{table}[ht]
\centering
\resizebox{\textwidth}{!}{
\begin{tabular}{|l|l|l|l|l|l|l|l|l|} 
\hline
\multicolumn{3}{|c|}{Round 1: Architecture}                                                                                                                                                                                            & \multicolumn{3}{c|}{Round 2: Learning Rate}                                                                                                                                                                                                           & \multicolumn{3}{c|}{Round 3: Noise Upper-bound}                                                                                                                                                                                                                                                       \\ 
\hline
\multicolumn{1}{|c|}{Architecture}  & \multicolumn{1}{c|}{{\begin{tabular}[c]{@{}c@{}}Train \\Loss (L2)\end{tabular}}} & \multicolumn{1}{c|}{{\begin{tabular}[c]{@{}c@{}}Test \\Loss (L2)\end{tabular}}} & \multicolumn{1}{c|}{{Learning Rate}} & \multicolumn{1}{c|}{{\begin{tabular}[c]{@{}c@{}}Train\\Loss (L2)\end{tabular}}} & \multicolumn{1}{c|}{{\begin{tabular}[c]{@{}c@{}}Test \\Loss (L2)\end{tabular}}} & \multicolumn{1}{c|}{{\begin{tabular}[c]{@{}c@{}}Noise \\Upper-bound\end{tabular}}} & \multicolumn{1}{c|}{{\begin{tabular}[c]{@{}c@{}}Train \\Loss (L2)\end{tabular}}} & \multicolumn{1}{c|}{{\begin{tabular}[c]{@{}c@{}}Test \\Loss (L2)\end{tabular}}}  \\ 
\multicolumn{1}{|c|}{(3 FC layers)} & \multicolumn{1}{c|}{}                                                                           & \multicolumn{1}{c|}{}                                                                          & \multicolumn{1}{c|}{}                               & \multicolumn{1}{c|}{}                                                                          & \multicolumn{1}{c|}{}                                                                          & \multicolumn{1}{c|}{}                                                                             & \multicolumn{1}{c|}{}                                                                           & \multicolumn{1}{c|}{}                                                                           \\ 
\hline
100, 100, 100                       & 0.130                                                                                           & 0.140                                                                                          & e-2                                                 & 0.163                                                                                          & 0.146                                                                                          & 0.9                                                                                               & 0.148                                                                                           & 0.169                                                                                           \\ 
\hline
200, 200, 200                       & 0.126                                                                                           & 0.161                                                                                          & e-3                                                 & 0.124                                                                                          & 0.139                                                                                          & 0.91                                                                                              & 0.140                                                                                           & 0.167                                                                                           \\ 
\hline
300, 300, 300                       & 0.125                                                                                           & 0.138                                                                                          & e-4                                                 & 0.182                                                                                          & 0.191                                                                                          & 0.92                                                                                              & 0.168                                                                                           & 0.181                                                                                           \\ 
\hline
400, 400, 400                       & 0.126                                                                                           & 0.142                                                                                          & e-5*                                                & 0.122                                                                                          & 0.122                                                                                          & 0.93                                                                                              & 0.139                                                                                           & 0.162                                                                                           \\ 
\hline
500, 500, 500                       & 0.128                                                                                           & 0.138                                                                                          & e-6                                                 & 0.127                                                                                          & 0.126                                                                                          & 0.94                                                                                              & 0.130                                                                                           & 0.143                                                                                           \\ 
\hline
600, 600, 600                       & 0.122                                                                                           & 0.147                                                                                          & e-7                                                 & 0.140                                                                                          & 0.131                                                                                          & 0.95*                                                                                             & 0.121                                                                                           & 0.137                                                                                           \\ 
\hline
700, 700, 700*                      & 0.127                                                                                           & 0.126                                                                                          & e-8                                                 & 0.163                                                                                          & 0.147                                                                                          & 0.96                                                                                              & 0.134                                                                                           & 0.166                                                                                           \\ 
\hline
800, 800, 800                       & 0.125                                                                                           & 0.157                                                                                          & e-9**                                               & 28.168                                                                                         & 28.293                                                                                         & 0.97                                                                                              & 0.142                                                                                           & 0.162                                                                                           \\ 
\hline
900, 900, 900                       & 0.126                                                                                           & 0.140                                                                                          & e-10**                                              & 38.406                                                                                         & 38.440                                                                                         & 0.98                                                                                              & 0.127                                                                                           & 0.163                                                                                           \\ 
\hline
1000, 1000, 1000                    & 0.122                                                                                           & 0.144                                                                                          & e-11**                                              & 38.757                                                                                         & 38.550                                                                                         & 0.99                                                                                              & 0.132                                                                                           & 0.149                                                                                           \\ 
\hline
                                    &                                                                                                 &                                                                                                & e-12**                                              & 38.341                                                                                         & 38.577                                                                                         & 1                                                                                                 & 0.138                                                                                           & 0.164                                                                                           \\
\hline
\end{tabular}
}
\caption{Dataset $B_s$ Hyperparameter Search}
\end{table}

\begin{table}[ht]
\resizebox{\textwidth}{!}{
\centering
\begin{tabular}{|l|l|l|l|l|l|l|l|} 
\hline
\multicolumn{1}{|c|}{{Train Loss}} & \multicolumn{1}{c|}{{Metric}}      & \multicolumn{6}{c|}{Training Optimizer}                                                                                                                                                                                                                       \\ 
\cline{3-8}
\multicolumn{1}{|c|}{}                            & \multicolumn{1}{c|}{}                             & \multicolumn{1}{c|}{{Adam (No Regularization)}} & \multicolumn{5}{c|}{AdamW (L2 Regularization)}                                                                                                                                               \\ 
% \hhline{|~~~-----|}
\multicolumn{1}{|c|}{}                            & \multicolumn{1}{c|}{}                             & \multicolumn{1}{c|}{}                                          & \multicolumn{1}{c|}{WD 1} & \multicolumn{1}{c|}{WD 0.1} & \multicolumn{1}{c|}{WD 0.01} & \multicolumn{1}{c|}{WD 0.001} & \multicolumn{1}{c|}{{\cellcolor[rgb]{0.514,0.753,0.514}}WD 0.0001}  \\ 
\hline
L1                                                & \% Final Dreamed Energy Below MF                  & 35\%                                                           & 20\%                      & 33\%                        & 39\%                         & 35\%                          & {\cellcolor[rgb]{0.514,0.753,0.514}}41\%                            \\ 
\hline
                                                  & \% Min Dreamed Energy Below MF                    & 39\%                                                           & 33\%                      & 48\%                        & 68\%                         & 52\%                          & {\cellcolor[rgb]{0.514,0.753,0.514}}59\%                            \\ 
\hline
                                                  & Lowest Final Dreamed Energy                       & -6.975506                                                      & -7.010591                 & -7.023497                   & -6.973194                    & -6.995163                     & {\cellcolor[rgb]{0.514,0.753,0.514}}-6.995487                       \\ 
\hline
                                                  & Lowest Min Dreamed Energy                         & -6.975506                                                      & -7.010591                 & -7.023497                   & -7.010591                    & -7.010591                     & {\cellcolor[rgb]{0.514,0.753,0.514}}-7.069694                       \\ 
\hline
                                                  & Training Trainset Loss                            & 0.168888                                                       & 0.177216                  & 0.131057                    & 0.122044                     & 0.151780                      & {\cellcolor[rgb]{0.514,0.753,0.514}}0.129265                        \\ 
\hline
                                                  & Training Testset Loss                             & 0.164494                                                       & 0.183575                  & 0.123223                    & 0.160083                     & 0.174067                      & {\cellcolor[rgb]{0.514,0.753,0.514}}0.155514                        \\ 
\hline
L2                                                & \% Final Dreamed Energy Below MF                  & 30\%                                                           & 23\%                      & 29\%                        & 27\%                         & 23\%                          & 27\%                                                                \\ 
\hline
                                                  & \% Min Dreamed Energy Below MF                    & 48\%                                                           & 34\%                      & 42\%                        & 35\%                         & 36\%                          & 43\%                                                                \\ 
\hline
                                                  & Lowest Final Dreamed Energy                       & -6.975895                                                      & -6.975895                 & -6.975895                   & -7.010591                    & -6.973194                     & -6.975895                                                           \\ 
\hline
                                                  & Lowest Min Dreamed Energy                         & -7.010164                                                      & -6.975895                 & -7.010591                   & -7.010591                    & -6.973194                     & -7.010591                                                           \\ 
\hline
                                                  & Training Trainset Loss                            & 0.135                                                          & 0.136                     & 0.125                       & 0.123                        & 0.113                         & 0.147                                                               \\ 
\hline
                                                  & Training Testset Loss                             & 0.135                                                          & 0.157                     & 0.129                       & 0.147                        & 0.157                         & 0.159                                                               \\
\hline
\end{tabular}
}
\caption{Dataset $B_s$ Loss Search}
\end{table}

\begin{table}[ht]
\centering
\resizebox{\textwidth}{!}{%
\begin{tabular}{|l|l|l|l|l|l|l|l|l|} 
\hline
\multicolumn{3}{|c|}{Round 1: Architecture}                                                                                                                                                                                           & \multicolumn{3}{c|}{Round 2: Learning Rate}                                                                                                                                                                                                            & \multicolumn{3}{c|}{Round 3: Noise Upper-bound}                                                                                                                                                                                                                                                       \\ 
\hline
\multicolumn{1}{|c|}{Architecture} & \multicolumn{1}{c|}{{\begin{tabular}[c]{@{}c@{}}Train \\Loss (L2)\end{tabular}}} & \multicolumn{1}{c|}{{\begin{tabular}[c]{@{}c@{}}Test \\Loss (L2)\end{tabular}}} & \multicolumn{1}{c|}{{Learning Rate}} & \multicolumn{1}{c|}{{\begin{tabular}[c]{@{}c@{}}Train \\Loss (L2)\end{tabular}}} & \multicolumn{1}{c|}{{\begin{tabular}[c]{@{}c@{}}Test \\Loss (L2)\end{tabular}}} & \multicolumn{1}{c|}{{\begin{tabular}[c]{@{}c@{}}Noise \\Upper-bound\end{tabular}}} & \multicolumn{1}{c|}{{\begin{tabular}[c]{@{}c@{}}Train \\Loss (L2)\end{tabular}}} & \multicolumn{1}{c|}{{\begin{tabular}[c]{@{}c@{}}Test \\Loss (L2)\end{tabular}}}  \\
(3 FC layers)                      & \multicolumn{1}{c|}{}                                                                           & \multicolumn{1}{c|}{}                                                                          & \multicolumn{1}{c|}{}                               & \multicolumn{1}{c|}{}                                                                           & \multicolumn{1}{c|}{}                                                                          & \multicolumn{1}{c|}{}                                                                             & \multicolumn{1}{c|}{}                                                                           & \multicolumn{1}{c|}{}                                                                           \\ 
\hline
100, 100, 100                      & 1.828                                                                                           & 1.742                                                                                          & e-2                                                 & 1.876                                                                                           & 1.867                                                                                          & 0.9                                                                                               & 1.666                                                                                           & 1.672                                                                                           \\ 
\hline
200, 200, 200                      & 1.768                                                                                           & 1.768                                                                                          & e-3                                                 & 1.767                                                                                           & 1.847                                                                                          & 0.91                                                                                              & 1.684                                                                                           & 1.681                                                                                           \\ 
\hline
300, 300, 300                      & 1.735                                                                                           & 1.785                                                                                          & e-4                                                 & 1.626                                                                                           & 1.804                                                                                          & 0.92                                                                                              & 1.669                                                                                           & 1.634                                                                                           \\ 
\hline
400, 400, 400                      & 1.696                                                                                           & 1.675                                                                                          & e-5                                                 & 1.624                                                                                           & 1.717                                                                                          & 0.93                                                                                              & 1.710                                                                                           & 1.539                                                                                           \\ 
\hline
500, 500, 500                      & 1.698                                                                                           & 1.880                                                                                          & e-6*                                                & 1.711                                                                                           & 1.535                                                                                          & 0.94                                                                                              & 1.695                                                                                           & 1.655                                                                                           \\ 
\hline
600, 600, 600                      & 1.662                                                                                           & 1.891                                                                                          & e-7                                                 & 1.825                                                                                           & 1.926                                                                                          & 0.95*                                                                                             & 1.711                                                                                           & 1.535                                                                                           \\ 
\hline
700, 700, 700*                     & 1.711                                                                                           & 1.535                                                                                          & e-8                                                 & 1.834                                                                                           & 2.134                                                                                          & 0.96                                                                                              & 1.669                                                                                           & 1.785                                                                                           \\ 
\hline
800, 800, 800                      & 1.659                                                                                           & 1.780                                                                                          & e-9**                                               & 10.067                                                                                          & 9.969                                                                                          & 0.97                                                                                              & 1.661                                                                                           & 1.914                                                                                           \\ 
\hline
900, 900, 900                      & 1.691                                                                                           & 1.628                                                                                          & e-10**                                              & 11.474                                                                                          & 11.549                                                                                         & 0.98                                                                                              & 1.668                                                                                           & 1.762                                                                                           \\ 
\hline
1000, 1000, 1000                   & 1.660                                                                                           & 1.779                                                                                          & e-11**                                              & 10.959                                                                                          & 11.568                                                                                         & 0.99                                                                                              & 1.667                                                                                           & 1.849                                                                                           \\ 
\hline
                                   &                                                                                                 &                                                                                                & e-12**                                              & 11.425                                                                                          & 10.849                                                                                         & 1                                                                                                 & 1.701                                                                                           & 1.750                                                                                           \\
\hline
\end{tabular}}
\caption{Dataset $C_s$ Hyperparameter Search}
\end{table}

\begin{table}[t]
\resizebox{\textwidth}{!}{%
\centering
\begin{tabular}{|l|l|l|l|l|l|l|l|} 
\hline
\multicolumn{1}{|c|}{{Train Loss}} & \multicolumn{1}{c|}{{Metric (n=100)}} & \multicolumn{6}{c|}{Training Optimizer}                                                                                                                                                                                                 \\ 
\cline{3-8}
\multicolumn{1}{|c|}{}                            & \multicolumn{1}{c|}{}                                & \multicolumn{1}{c|}{{Adam (No Regularization)}} & \multicolumn{5}{c|}{AdamW (L2 Regularization)}                                                                                                                         \\ 
\cline{4-8}
\multicolumn{1}{|c|}{}                            & \multicolumn{1}{c|}{}                                & \multicolumn{1}{c|}{}                                          & \multicolumn{1}{c|}{WD 1} & \multicolumn{1}{c|}{WD 0.1} & \multicolumn{1}{c|}{WD 0.01} & \multicolumn{1}{c|}{WD 0.001} & \multicolumn{1}{c|}{WD 0.0001}                \\ 
\hline
L1                                                & \% Final Dreamed Energy Below MF                     & 4\%                                                            & 5\%                       & 1\%                         & 3\%                          & 2\%                           & 5\%                                           \\ 
\hline
                                                  & \% Min Dreamed Energy Below MF                       & 12\%                                                           & 19\%                      & 11\%                        & 7\%                          & 15\%                          & 18\%                                          \\ 
\hline
                                                  & Lowest Final Dreamed Energy                          & -6.96606                                                       & -6.96938                  & -6.92135                    & -6.97532                     & -6.95646                      & -6.96508                                      \\ 
\hline
                                                  & Lowest Min Dreamed Energy                            & -6.96577                                                       & -7.03760                  & -7.00561                    & -6.97531                     & -6.96430                      & -6.97183                                      \\ 
\hline
                                                  & Training Trainset Loss                               & 1.7084291                                                      & 1.7229033                 & 1.6923031                   & 1.6598003                    & 1.686494                      & 1.7273679                                     \\ 
\hline
                                                  & Training Testset Loss                                & 1.7635838                                                      & 1.7890829                 & 1.8463715                   & 1.8677194                    & 1.9780074                     & 1.7214417                                     \\ 
\hline
L2                                                & \% Final Dreamed Energy Below MF                     & 5\%                                                            & 3\%                       & 4\%                         & 9\%                          & 7\%                           & {\cellcolor[rgb]{0.514,0.753,0.514}}4\%       \\ 
\hline
                                                  & \% Min Dreamed Energy Below MF                       & 18\%                                                           & 14\%                      & 21\%                        & 23\%                         & 18\%                          & {\cellcolor[rgb]{0.514,0.753,0.514}}6\%       \\ 
\hline
                                                  & Lowest Final Dreamed Energy                          & -7.01137                                                       & -6.98527                  & -6.95640                    & -6.97303                     & -6.97490                      & {\cellcolor[rgb]{0.514,0.753,0.514}}-7.05958  \\ 
\hline
                                                  & Lowest Min Dreamed Energy                            & -7.02930                                                       & -6.99054                  & -6.97814                    & -6.98117                     & -7.00088                      & {\cellcolor[rgb]{0.514,0.753,0.514}}-7.07089  \\ 
\hline
                                                  & Training Trainset Loss                               & 1.691                                                          & 1.670                     & 1.695                       & 1.682                        & 1.685                         & {\cellcolor[rgb]{0.514,0.753,0.514}}1.657     \\ 
\hline
                                                  & Training Testset Loss                                & 1.650                                                          & 1.879                     & 1.689                       & 1.631                        & 1.682                         & {\cellcolor[rgb]{0.514,0.753,0.514}}1.836     \\
\hline
\end{tabular}}
\caption{Dataset $C_s$ Loss Search}
\end{table}